\documentclass[11pt]{article}
\pdfoutput=1
\usepackage[a4paper,margin=3cm,bmargin=4.5cm]{geometry}
\usepackage{graphicx}
\usepackage{amssymb,amsmath}
\usepackage[table]{xcolor} %for \rowcolors
\usepackage{slashed}
\usepackage{hyperref}
\usepackage{booktabs}%for \toprule
\usepackage{colortbl} %for \rowcolor
\usepackage{braket}
\usepackage{subcaption}
\usepackage{slashed}
\newcommand{\Slash}[1]{{\ooalign{\hfil#1\hfil\crcr\raise.167ex\hbox{/}}}}
\usepackage[separate-uncertainty=true]{siunitx}
\usepackage{cite}
\usepackage{cancel}
\usepackage[utf8]{inputenc}
\usepackage{floatpag}

\usepackage[compat=1.1.0]{tikz-feynman}
\tikzfeynmanset{
  every edge={thick},
  warn luatex=false, % to suppress warning
}

\makeatletter
\def\tikzfeynman@luatex@required@path{}
\def\tikzfeynman@luatex@required@key{}
\makeatother

\hypersetup{
  colorlinks = true,
  linkcolor = blue,
  citecolor = blue,
}

\let\tilde\widetilde

\def\beq#1\eeq{\begin{align}#1\end{align}}

\newcommand\w[1]{_{\mathrm{#1}}}

\newcommand{\eg}{{\em e.g.}}
\newcommand{\ie}{{\em i.e.}}

\newcommand\unit[1]{\,\mathrm{#1}}
\newcommand\GeV{\unit{GeV}}
\newcommand\TeV{\unit{TeV}}

\newcommand\iab{\unit{ab^{-1}}}
\newcommand\ifb{\unit{fb^{-1}}}

\newcommand\amu[1][\relax]{\ifx#1\relax{a_\mu}\else{a_\mu^{\mathrm{#1}}}\fi}
\newcommand\sel{\tilde e}
\newcommand\smu{\tilde\mu}
\newcommand\stau{\tilde\tau}
\newcommand\smuL{\tilde\mu\w L}

\newcommand\neut  [1][\relax]{{\tilde\chi^0_{#1}}}
\newcommand\charP [1][\relax]{{\tilde\chi^+_{#1}}}
\newcommand\charM [1][\relax]{{\tilde\chi^-_{#1}}}
\newcommand\charPM[1][\relax]{{\tilde\chi^\pm_{#1}}}

\newcommand\package[2][\relax]{\texttt{#2}\ifx#1\relax\relax\else~\texttt{#1}\fi}

\renewcommand{\baselinestretch}{1.1} %space between the lines

\begin{document}

\begin{titlepage}
\setcounter{page}{0} % to avoid duplication of P.1
\begin{flushright}
IPMU21--0027\\ % \TODO{fill later!}
KEK--TH--2319 % \TODO{fill later!}
\end{flushright}
\vskip 1.5cm
\begin{center}
  {\Large \bf
  Supersymmetric Interpretation of 
  the Muon $\boldsymbol{g-2}$
  Anomaly
  }
\vskip 1.5cm
{
  Motoi Endo,$^{(a,b,c)}$
  Koichi Hamaguchi,$^{(c,d)}$
  Sho Iwamoto,$^{(e)}$
  and
  Teppei Kitahara$^{(f,g)}$
}

\vspace{1.5em}

\begingroup\small\itshape
\begin{tabbing}
$^{(a)}$ \!\!\= KEK Theory Center, IPNS, KEK, Tsukuba, Ibaraki 305--0801, Japan
\\[0.3em]
$^{(b)}$ \> The Graduate University of Advanced Studies (Sokendai), Tsukuba, Ibaraki 305--0801, Japan
\\[0.3em]
$^{(c)}$ \> Kavli IPMU (WPI), UTIAS, The University of Tokyo, Kashiwa, Chiba 277--8583, Japan
\\[0.3em]
$^{(d)}$ \> Department of Physics, The University of Tokyo, Bunkyo-ku, Tokyo 113--0033, Japan
\\[0.3em]
$^{(e)}$ \> ELTE E\a"otv\a"os Lor\a'and University, P\a'azm\a'any P\a'eter s\a'et\a'any 1/A, Budapest H-1117, Hungary\\[0.3em]
$^{(f)}$ \> Institute for Advanced Research, Nagoya University, Nagoya 464--8601, Japan
\\[0.3em]
$^{(g)}$ \> Kobayashi-Maskawa Institute for the Origin of Particles and the Universe, \\
         \> Nagoya University,  Nagoya 464--8602, Japan
\end{tabbing}
\endgroup

\vspace{1cm}

\abstract{
\noindent 
The Fermilab Muon $g-2$ collaboration recently announced 
the first 
result of measurement 
of the muon anomalous magnetic moment ($g-2$),
which confirmed 
the previous result at the Brookhaven National Laboratory
and thus the discrepancy with its Standard Model prediction.
We revisit low-scale supersymmetric models that are naturally capable to solve the muon $g-2$ anomaly, focusing on two distinct scenarios: 
chargino-contribution dominated and pure-bino-contribution dominated scenarios.
It is shown that
the slepton pair-production searches have excluded broad parameter spaces for both two scenarios, but
they are not closed yet.
For the chargino-dominated scenario, the models with $m_{\tilde{\mu}_{\rm L}}\gtrsim m_{\tilde{\chi}^{\pm}_1}$ are still widely allowed.
For the bino-dominated scenario, we find that, although slightly non-trivial, the region with low $\tan \beta$ with heavy higgsinos is preferred.
In the case of universal slepton masses, 
the low mass regions with $m_{\tilde{\mu}}\lesssim 230\GeV$ can explain the $g-2$ anomaly while satisfying the LHC constraints.
Furthermore, we checked that the stau-bino coannihilation works properly
to realize the bino thermal relic dark matter. 
We also investigate heavy staus case for the bino-dominated scenario, where 
the parameter region that can explain the muon $g-2$ anomaly is stretched to $m_{\tilde{\mu}}\lesssim 1.3\TeV$.
}

\vspace{2em}
% official JHEP keywords from https://jhep.sissa.it/jhep/help/keywordsList.jsp
\end{center}
\end{titlepage}

\renewcommand{\thefootnote}{\#\arabic{footnote}}
\setcounter{footnote}{0}

%%%%%%%%%%%%%%%%%%%%%%%%%
% Contents
%%%%%%%%%%%%%%%%%%%%%%%%%
\begingroup
\renewcommand{\baselinestretch}{1.3}

\hrule
\tableofcontents
\vskip .2in
\hrule
\vskip .45in
\endgroup

%%%%%%%%%%%%%%%%%%%%%%%%%%%%%%%%%%%%%%%%%%%%%%%%%%
\section{Introduction}
\label{sec:introduction}
%%%%%%%%%%%%%%%%%%%%%%%%%%%%%%%%%%%%%%%%%%%%%%%%%%

The success of the Standard Model (SM) has been confirmed by the Higgs boson discovery,
the Higgs coupling measurements,
recent results from the Large Hadron Collider (LHC), 
and a number of low-energy precision measurements in the quark and lepton flavor physics.
However, physics beyond the SM (BSM) is definitely required, for example, in order to explain the dark matter.

The muon anomalous magnetic moment [the muon $g-2$; $a_{\mu} \equiv (g_{\mu}-2)/2$] may be a hint to construct BSM scenarios, for there has laid $3.7\sigma$-level discrepancy between its SM prediction~\cite{Aoyama:2020ynm},%
\footnote{
Recent development of the lattice calculation including QED and isospin-breaking corrections  \cite{Borsanyi:2020mff}
has revealed that 
the leading-order hadronic vacuum polarization contributions disagree with those calculated by
the data-driven approach at $2.3\sigma$ level \cite{Aoyama:2020ynm,Colangelo:2018mtw,Hoferichter:2019mqg,Davier:2019can,Keshavarzi:2019abf} and the resultant $\Delta a_{\mu} $ is zero-consistent within $2\sigma$ level.
In order to explain the $2.3\sigma$ discrepancy while  saving the global fit of the electroweak sector,
the cross section $\sigma(e^+ e^- \to\text{hadrons})$ below $1\GeV$ region has to be changed
\cite{Crivellin:2020zul,Keshavarzi:2020bfy,Malaescu:2020zuc}.
A dedicated study for such the energy region is given in Ref.~\cite{Colangelo:2020lcg}.}
\begin{equation}
 \amu[\text{SM}] = \left( 11\,659\,181.0 \pm 4.3 \right) \times 10^{-10}\,,
\end{equation}
and the value measured at the Brookhaven National Laboratory in 1997--2001~\cite{Bennett:2002jb,Bennett:2004pv,Bennett:2006fi},\footnote{%
This value is calculated with the latest value of the muon-to-proton magnetic ratio~\cite{Aoyama:2020ynm,CODATA2018}.}
\begin{equation}
 a_{\mu}^{\rm BNL}  = \left(11\,659\,208.9 \pm 5.4_{\rm stat} \pm 3.3_{\rm sys} \right) \times 10^{-10}\,.
\label{eq:amuBNL}
\end{equation}

The effort on experimental reconfirmation of this anomaly has been made 
by the Fermilab Muon $g-2$ collaboration \cite{Grange:2015fou} 
and the J-PARC muon $g-2$/EDM collaboration \cite{Mibe:2011zz, Abe:2019thb}.
These on-going experiments aim to reduce the experimental error 
at least  by a factor of four
compared to $ a_{\mu}^{\rm BNL} $.
Very recently, the Fermilab Muon $g-2$ collaboration 
presented the first 
result on the measurement~\cite{g-2Seminar20210407,Abi:2021gix},
\begin{equation}
 \amu[\text{FNAL; 2021Apr.}] =
\left(11\,659\,204.0 \pm 5.4 \right) \times 10^{-10}\,,
\label{eq:amuFNAL}
\end{equation}
corresponding to $3.3\sigma$ level. 
Together with the BNL measurement, 
the average value of the $\amu$ measurements is given by \cite{g-2Seminar20210407,Abi:2021gix}
\begin{equation}
 \amu[\text{BNL+FNAL}] =
\left(11\,659\,206.1 \pm 4.1 \right) \times 10^{-10}\,.
\end{equation}
Now the discrepancy between the experimental and theoretical values amounts to
\begin{equation}
 \Delta a_{\mu}
 \equiv  \amu[\text{BNL+FNAL}]  - \amu[SM]
 =      \left( 25.1\pm 5.9\right) \times 10^{-10}\,,
 \label{eq:Deltaamu}
\end{equation}
whose significance is equivalent to $4.2\sigma$ level,
and the muon $g-2$ anomaly is reconfirmed.\footnote{Very recently, the hadronic light-by-light contribution was analyzed in the first principle by the lattice QCD calculation~\cite{Chao:2021tvp}. The significance of the anomaly could decrease slightly.}

This discrepancy is as large as the SM electroweak contribution to the muon $g-2$,
$a_\mu(\text{EW})= (15.4\pm 0.1)\times 10^{-10}$ \cite{Aoyama:2020ynm},
which implies that BSM physics around the electroweak scale 
may be responsible for it.
Low-energy supersymmetry (SUSY) is one of such solutions.
Its contribution to the muon $g-2$, which we denote by $\amu[SUSY]$, can naturally explain the discrepancy because it is amplified by  model-specific parameters as we shall see~\cite{Lopez:1993vi,Chattopadhyay:1995ae,Moroi:1995yh}.
One can also benefit from the merits of SUSY models, such as the explanation of the gauge hierarchy problem, the gauge coupling unification,
and the existence of the dark matter candidate as the  lightest SUSY particle (LSP); this makes SUSY more attractive.

The SUSY contributions to the muon $g-2$
can be sizable when 
at least {\it three} SUSY multiplets are as light as $\mathcal{O}(100)\GeV$.
They are classified into four types: ``WHL'', ``BHL'', ``BHR'', and ``BLR'',
where W, B, H, L, and R stand for wino, bino, higgsino, left-handed and right-handed smuons, respectively.
Under the mass-insertion approximation,
these four types are given as  \cite{Moroi:1995yh}\footnote{Note that the exact analytic formulae are used in our analysis instead of relying on this approximation. See discussion below.}
\begin{align}
  \amu[WHL]
    &=\frac{\alpha_2}{4\pi} \frac{m_{\mu}^2}{M_2\mu} \tan\beta\cdot
    f_C\left(\frac{M_2^2}{m_{\tilde{\nu}_{\mu}}^2}, \frac{\mu^2}{m_{\tilde{\nu}_{\mu}}^2} \right) -\frac{\alpha_2}{8\pi} \frac{m_{\mu}^2}{M_2\mu} \tan\beta\cdot
    f_N\left(\frac{M_2^2}{m_{\tilde{\mu}\w L}^2}, \frac{\mu^2}{m_{\tilde{\mu}\w L}^2} \right)\,,
    \label{eq:WHL} \\
  \amu[BHL]
  &= \frac{\alpha_Y}{8\pi} \frac{m_{\mu}^2}{M_1\mu} \tan\beta\cdot
    f_N\left(\frac{M_1^2}{m_{\tilde{\mu}\w L}^2}, \frac{\mu^2}{m_{\tilde{\mu}\w L}^2} \right)\,,
    \label{eq:BHL} \\
  \amu[BHR]
  &= - \frac{\alpha_Y}{4\pi} \frac{m_{\mu}^2}{M_1\mu} \tan\beta\cdot
    f_N\left(\frac{M_1^2}{m_{\tilde{\mu}\w R}^2}, \frac{\mu^2}{m_{\tilde{\mu}\w R}^2} \right)\,,
    \label{eq:BHR} \\
  \amu[BLR]
  &= \frac{\alpha_Y}{4\pi} \frac{m_{\mu}^2M_1\mu}{m_{\tilde{\mu}\w L}^2m_{\tilde{\mu}\w R}^2}
    \tan\beta\cdot
    f_N\left(\frac{m_{\tilde{\mu}\w L}^2}{M_1^2}, \frac{m_{\tilde{\mu}\w R}^2}{M_1^2} \right)\,,
    \label{eq:BLR} 
\end{align}
where $M_1$ ($M_2$) is the bino (wino) soft mass,
$\mu$ is the higgsino mass parameter,
$\tan\beta=v_u/v_d$ is the ratio of the vacuum expectation values of the up- and down-type Higgs,
and $m_{\tilde{\mu}\w {L/R}}$ and $m_{\tilde{\nu}_{\mu}}$ are the masses of the left/right-handed smuon and the muon sneutrino, respectively.
The loop functions are given by
\begin{align}
    \label{eq:loop-aprox}
    f_C(x,y)
    &=xy\left[
      \frac{5-3(x+y)+xy}{(x-1)^2(y-1)^2} - \frac{2\ln x}{(x-y)(x-1)^3}+\frac{2\ln y}{(x-y)(y-1)^3}
      \right]\,,
      \\
    f_N(x,y)
    &= xy\left[
      \frac{-3+x+y+xy}{(x-1)^2(y-1)^2} + \frac{2x\ln x}{(x-y)(x-1)^3}-\frac{2y\ln y}{(x-y)(y-1)^3}
      \right]\,,
\end{align}
which satisfy $0\le f_{C/N}(x,y)\le 1$, $f_C(1,1) = 1/2$ and $f_N(1,1) = 1/6$.
An important fact is that 
the first three contributions, $\amu[WHL]$,  $\amu[BHL]$,  and $\amu[BHR]$, are enhanced by $\tan \beta$ but suppressed as $\mu$ increases.
On the other hand, the last one,  $\amu[BLR]$, is enhanced by $\mu \tan \beta$.
This difference provides us with two completely distinct scenarios.

In this paper, we revisit two branches of the SUSY scenarios proposed to explain the muon $g-2$ anomaly:\footnote{For recent studies of the muon $g-2$ motivated SUSY models based on the LHC Run~2 results, see, \eg, Refs.~\cite{Zhu:2016ncq,Choudhury:2017fuu,Yanagida:2017dao,Endo:2017zrj,Hagiwara:2017lse,Chakraborti:2017vxz,Choudhury:2017acn,Ajaib:2017zba,Belyaev:2018vkl,Bhattacharyya:2018inr,Abel:2018ekz,Cao:2018rix,Dutta:2018fge,Cox:2018vsv,Tran:2018kxv,Ibe:2019jbx,Badziak:2019gaf,Abdughani:2019wai,Kpatcha:2019pve,Yanagida:2020jzy,Han:2020exx,Chakraborti:2020vjp,Nagai:2020xbq,Chakraborti:2021kkr}. } one in which the chargino contribution, $\amu[WHL]$, is dominant, and the other in which the pure-bino contribution, $\amu[BLR]$, is dominant.\footnote{The other contributions, $\amu[BHL]$ and $\amu[BHR]$, can also be dominant. See, \eg,~Ref.~\cite{Endo:2017zrj}.}
This work focuses on the minimal setups, 
where only three or four SUSY multiplets are light to have sizable $\amu[WHL]$ or $\amu[BLR]$, 
while the other
irrelevant particles are decoupled.
Although this approach lacks specific SUSY-breaking models and does not cover the whole possible Minimal Supersymmetric Standard Model (MSSM) models,
it allows us to clarify the relevant MSSM parameters and tell which LHC (and other) constraints are relevant, compared to studies with scans over the whole parameter space.\footnote{%
This strategy has been taken 
in, \eg, the following works: Refs.~\cite{Hagiwara:2017lse,Chakraborti:2017vxz,Endo:2020mqz} with particular focus on WHL,
Refs.~\cite{Endo:2017zrj,Abdughani:2019wai} on BHL and BHR,
and
Refs.~\cite{Abdughani:2019wai,Chakraborti:2020vjp,Chakraborti:2021kkr} considering the combination of the contributions.}

This paper is organized as follows.
%\begin{itemize}
%    \item 
    In Sec.~\ref{sec:chargino}, we revisit the parameter regions in which $\amu[SUSY]\approx\amu[WHL]$, based on our previous study~\cite{Endo:2020mqz}. Compared with Ref.~\cite{Endo:2020mqz}, the new $\amu$ measurements~\cite{Abi:2021gix}, the new theory combination~\cite{Aoyama:2020ynm}, and new results from the LHC~\cite{Aad:2020qnn,Sirunyan:2020eab} are implemented. 
    %
%    \item 
    In Sec.~\ref{sec:bino}, we revisit 
    the bino-contribution to the muon $g-2$, $\amu[BLR]$, taking into account of the latest LHC constraints as well as the vacuum meta-stability. A related analysis was performed in our previous study~\cite{Endo:2013lva}. Interestingly, a wider region of the parameter space is allowed for smaller $\tan\beta$, rather than the large $\tan\beta$ case. We study the cases with and without flavor universal slepton masses. We also discuss an implication for the dark matter abundance with the bino-slepton coannihilation.
%\end{itemize}
Section~\ref{sec:conclusion} is devoted to conclusions and discussion.

%%%%%%%%%%%%%%%%%%%%%%%%%%%%%%%%%%%%%%%%%%%%%%%%%
\section{Chargino contributions}
\label{sec:chargino}
%%%%%%%%%%%%%%%%%%%%%%%%%%%%%%%%%%%%%%%%%%%%%%%%%%

%
In this section,  we consider the parameter regions where the chargino contribution to the muon $g-2$ is dominant, based on our previous study~\cite{Endo:2020mqz}.

%%%%%%%%%%
\subsection{Setup and Result}

The setup in this section is as follows:
\begin{itemize}
\item
Among SUSY particles, neutralinos $\neut[i]$, charginos $\charPM[j]$, and left-handed sleptons $\tilde l\w L$, $\tilde\nu$ are within the LHC reach, \ie, with masses of $\lesssim1\TeV$.
Here, $\tilde{l}$ denotes $\tilde{e},\tilde{\mu}$, and $\tilde{\tau}$.
\item The right-handed sleptons $\tilde l\w R$ are heavy, so that $\amu[BHR]$ and $\amu[BLR]$ are suppressed and our analysis is simplified. The scalar trilinear terms $(A_e)_{ij}$ are neglected for simplicity.
\item The soft masses of the left-handed sleptons are flavor universal and diagonal.
\item All the colored SUSY particles (gluino and squarks) are decoupled.
This assumption makes the LHC constraints more conservative, while their contribution to the muon $g-2$ is negligibly small even if they are light because it arises at the two-loop level.
In addition, heavy colored SUSY particles are motivated by the mass of the SM-like Higgs boson as well as by LHC constraints.
\item The heavy Higgs bosons, whose contributions to $\amu[SUSY]$ are also negligible, are decoupled as well.
\end{itemize}
Then, the following five model parameters are left relevant:
\begin{equation}
    M_1,\quad  M_2, \quad \mu, \quad  m^2\w L, \quad \tan\beta,
    \label{eq:parameters}
\end{equation}
where $m\w L$ stands for the universal soft mass for the left-handed sleptons,
and as in Ref.~\cite{Endo:2020mqz},
the following four subspace of the parameters are analyzed:\footnote{%
  We here require that the LSP is the bino-like neutralino with ignorance of the dark matter overabundance. The relic density may be reduced to be consistent with the observed one by an entropy production or considering more generic parameter space of the SUSY models (cf.~\cite{Chakraborti:2021kkr}).
 Meanwhile, Ref.~\cite{Iwamoto:2021aaf} focuses on a similar scenario in which $\amu[WHL]$ explains the anomaly with $M_1>M_2,\mu$, \ie, a wino-higgsino mixed neutralino as the LSP.
 }
\begin{alignat}{3}
 \text{(A)}\;
 & M_1=\frac{1}{2}M_2\,, &\quad &\mu=M_2\,,&\quad  &\tan\beta=40\,,
 \\
 \text{(B)}\;
 & M_1=\frac{1}{2}M_2\,, &\;&\mu=2M_2\,,&\; &\tan\beta=40\,,
 \\
 \text{(C)}\;
 & m_{\tilde{\chi}^0_1} = 100\,\textrm{GeV}\,, &\; &\mu=M_2\,,&\; &\tan\beta=40\,,
 \\
 \text{(D)}\;
 & m_{\tilde{\chi}^0_1} = 100\,\textrm{GeV}\,, & \; &\mu=2M_2\,, &\;&\tan\beta=40\,.
\label{eq:parameterspace}
\end{alignat}
The analyses is done in the same manner as in Ref.~\cite{Endo:2020mqz}.
In particular,
\package[1.5a]{SDECAY} \cite{Muhlleitner:2003vg,Djouadi:2006bz} and
\package[1.5.0]{GM2Calc} \cite{Athron:2015rva}
are utilized to calculate the decay rates and $\amu[SUSY]$, respectively, and
the right-handed slepton mass parameter $m\w R$ is taken to be $m\w R = 3\TeV$.

\begin{figure}[p]
 \centering
 \renewcommand\thesubfigure{\Alph{subfigure}}
%%%%%
  \begin{subfigure}[b]{0.49\textwidth}
 \includegraphics[width=\textwidth]{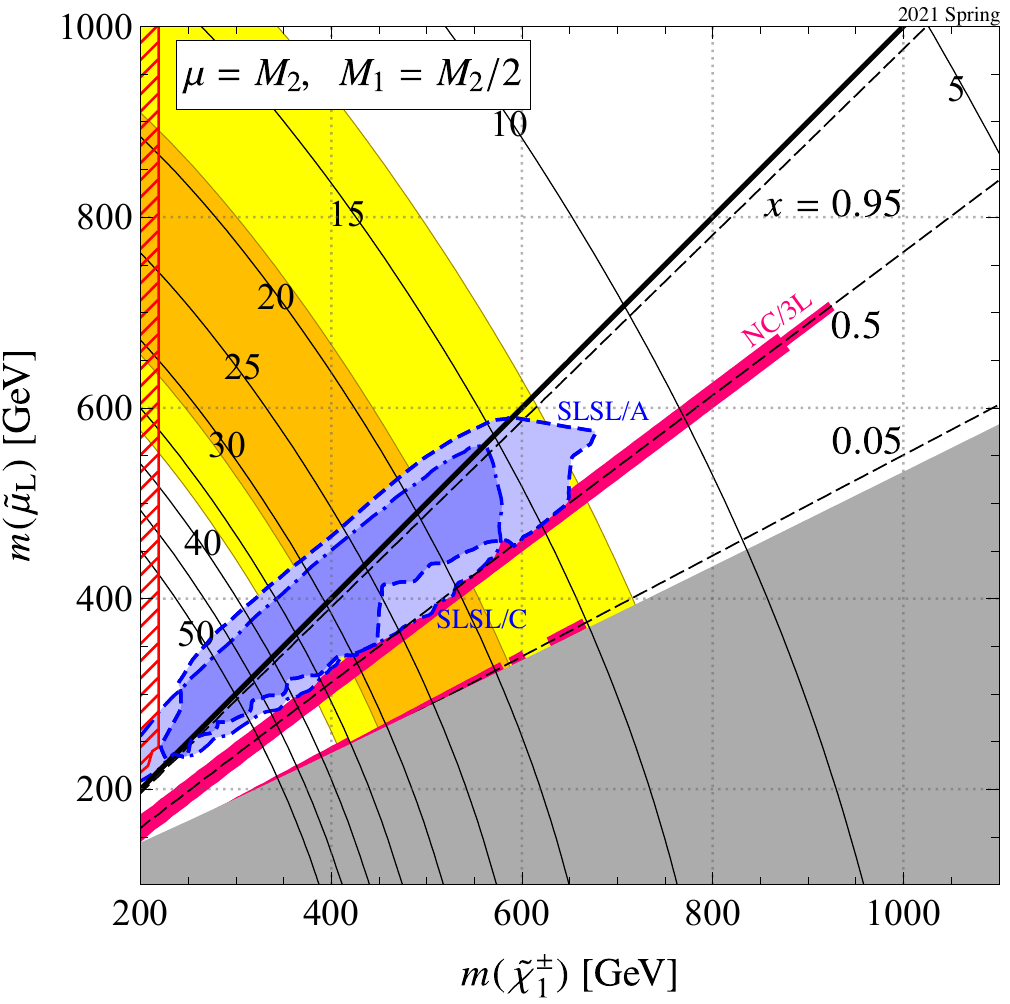}
\caption{$\mu = M_2$, $M_1 = M_2/2$}
 \vspace{.2cm}
 \end{subfigure}
 %%%
   \begin{subfigure}[b]{0.49\textwidth}
 \includegraphics[width=\textwidth]{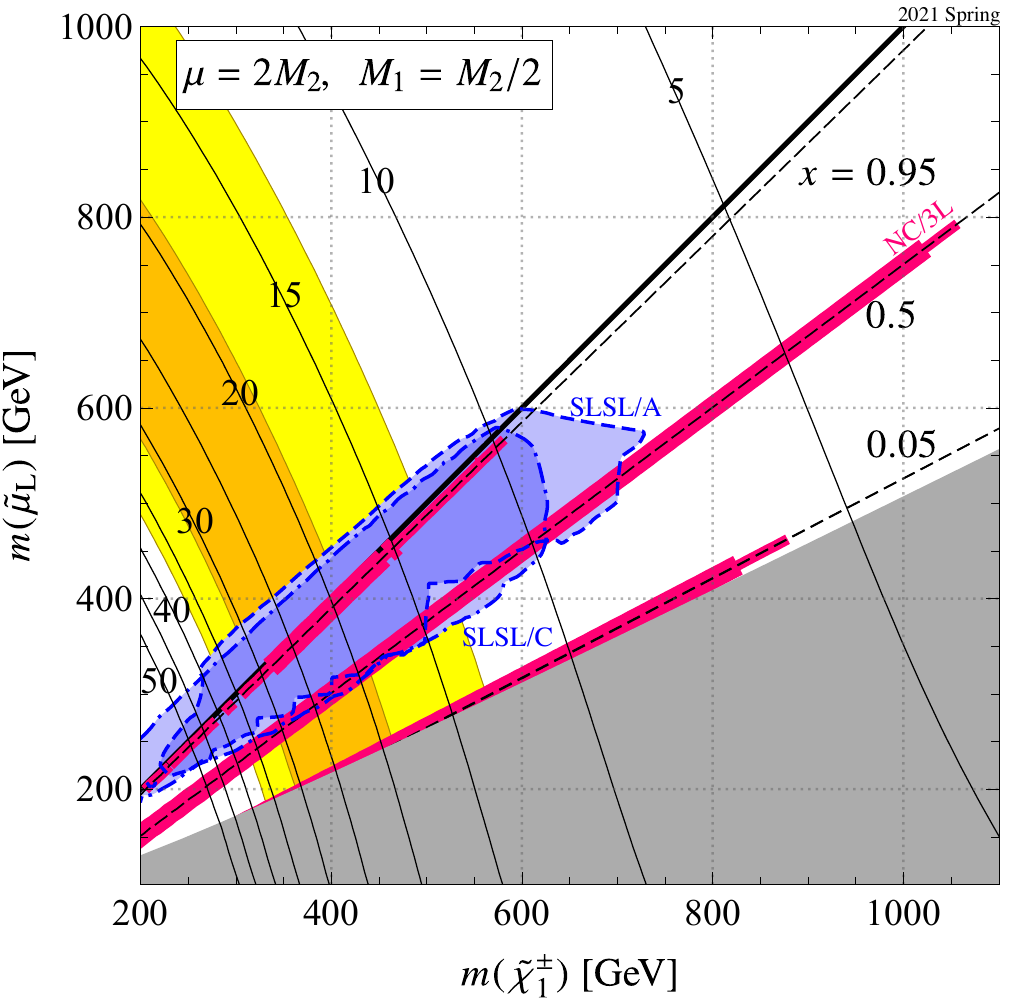}
\caption{$\mu = 2 M_2$, $M_1 = M_2/2$}
 \vspace{.2cm}
 \end{subfigure}
 %%%
   \begin{subfigure}[b]{0.49\textwidth}
 \includegraphics[width=\textwidth]{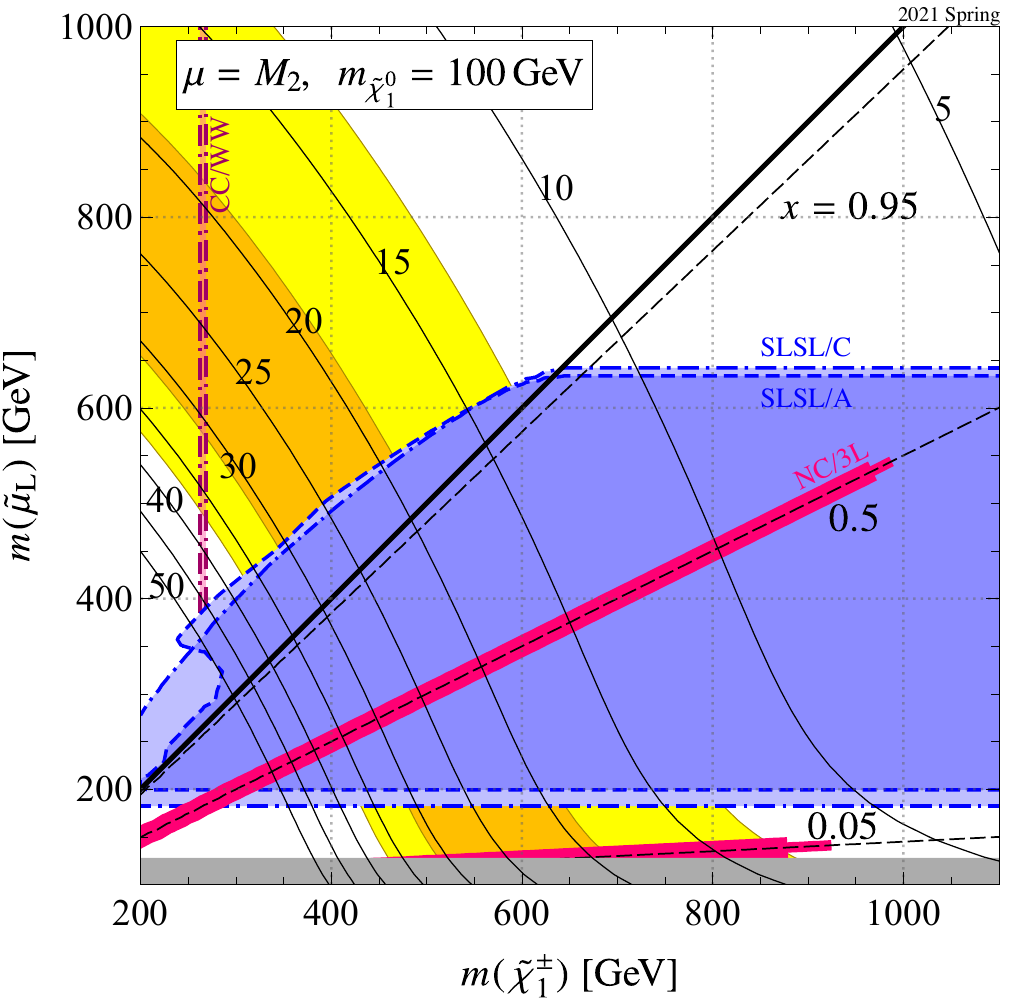}
\caption{$\mu = M_2$, $m_{\tilde{\chi}^0_1}=100$\,GeV}
 \end{subfigure}
 %%%
   \begin{subfigure}[b]{0.49\textwidth}
  \includegraphics[width=\textwidth]{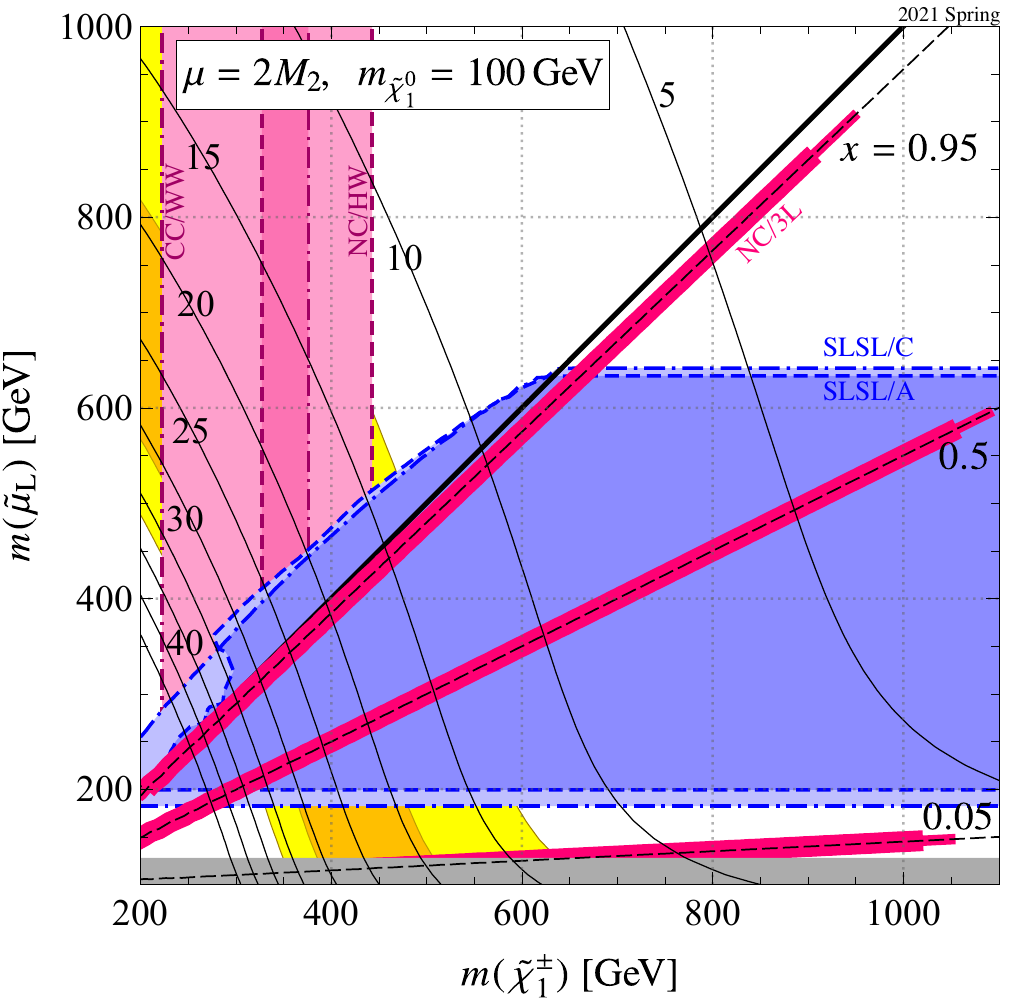}
\caption{$\mu = 2 M_2$, $m_{\tilde{\chi}^0_1}=100$\,GeV}
 \end{subfigure}
 %%%%%
  \caption{\label{fig:chargino}%
  The 2021 Spring summary of the chargino-dominated SUSY scenario for the muon $g-2$ anomaly.
  Four benchmark parameter planes are considered, where the WHL contribution is sizable and $\amu[SUSY]$ explains the anomaly at the $1\sigma$ ($2\sigma$) level in the orange-filled (yellow-filled) regions; $\amu[SUSY]\times10^{10}$ is shown by the black contours (but up to 50).
  The thick black line corresponds to $m_{\tilde{\mu}\w L} = m_{\charPM[1]}$.
  The gray-filled region, where the LSP is $\tilde\nu$, and the red-hatched region in (A), which corresponds to a compressed spectrum, are not studied.
  The red-filled and blue-filled regions are excluded by the LHC experiment~\cite{Aad:2019vnb,Aad:2019vvf,Sirunyan:2020eab}.
  We also analyzed the results of Refs.~\cite{Sirunyan:2017lae,Aaboud:2018jiw} but only on the model points with $x=0.05$, $0.5$, and $0.95$ [see Eq.~\eqref{eq:x}]; the excluded ranges are shown by the magenta bars.
  Detailed description of the LHC constraints is provided in our previous work~\cite{Endo:2020mqz}.
  }
\end{figure}

The results are shown in Fig.~\ref{fig:chargino} under the axes being the physical masses of the lighter chargino $\charPM[1]$ and the left-handed smuon $\tilde\mu\w L$.
The SUSY contribution to the muon $g-2$ is shown by the black solid contours in terms of $\amu[SUSY]\times10^{10}$. 
The parameter spaces where $\amu[SUSY]$ solves the discrepancy $\Delta\amu=(25.1\pm 5.9)\times10^{-10}$ at the $1\sigma$ ($2\sigma$) level are shown by the orange-filled (yellow-filled) regions. 
The LHC Run~2 constraints are shown by blue-filled regions, red-filled regions, and magenta lines at the 95\% confidence level; they are described in the next subsection.

We find that, as far as considering the parameter space in Fig.~\ref{fig:chargino}, models with $m_{\tilde{\mu}\w L} < m_{\tilde{\chi}^{\pm}_1}$ are strongly disfavored as a solution to the muon $g-2$ anomaly.
Meanwhile, for models with $m_{\tilde{\mu}\w L} > m_{\tilde{\chi}^{\pm}_1}$, LHC constraints are not critical yet; those models may explain the muon $g-2$ anomaly and are to be searched for at the future LHC runs.
In Table~\ref{tab:BMP-c}, we show the mass spectra and the SUSY contribution to the muon $g-2$ ($\amu[SUSY]$) for some viable benchmark points in Fig.~\ref{fig:neutralinoDM}.
The production cross sections and the branching fractions of the SUSY particles for those points can be found in Ref.~\cite{Endo:2020mqz}.

%%%%%%%%%%%%%%%%%%%%%%%%%%%%%%%%%%%%%%%%%%%%%%%%%%
%%%%%%%%%%%%%%%%%%%%%%%%%%%%%%%
\begin{table}[t]
\centering
\newcommand{\bhline}[1]{\noalign{\hrule height #1}}
\renewcommand{\arraystretch}{1.5}
\rowcolors{2}{gray!15}{white}
\addtolength{\tabcolsep}{5pt} % add space between columns
\caption{\label{tab:BMP-c} 
Benchmark points for the chargino-contribution dominated scenario.
The mass parameters are in units of GeV.
}
  \begin{tabular}{c ccc} 
  %\toprule
  \bhline{1 pt}
  \rowcolor{white}
  & WHL1 & WHL2 & WHL3 \\  \hline 
 $M_1$ & 200 & 200 & 102   \\  
 $M_2$ & 400 & 400 & 400   \\  
 $\mu$ & 400 & 800 & 400   \\  
 $m\w L$ & 600 & 600 & 600  \\ 
 $\tan \beta$ & 40 & 40 & 40 \\   
  \hline 
 $m_{\tilde{\mu}_1}, m_{\tilde{\tau}_1}$ & 602 & 602 & 602  \\ 
  $ m_{\tilde{\nu}_{\mu,\tau}}$ & 597 & 597 & 597  \\ 
 $m_{\tilde{\chi}_1^0}$ & 196 & 199 & 100 \\ 
 $m_{\tilde{\chi}_2^0}$ & 347 & 394 & 346  \\
  $m_{\tilde{\chi}_3^0}$ & 405 & 803 & 406  \\
   $m_{\tilde{\chi}_4^0}$ & 462 & 810 & 461  \\
   $ m_{\tilde{\chi}_1^{\pm}}$ & 346 & 394 & 346  \\
      $ m_{\tilde{\chi}_2^{\pm}}$ & 462 & 811 & 462  \\
 \hline 
 $\amu[SUSY]\times 10^{10}$ & 24 & 15 & 24  \\ 
\bhline{1 pt}
%\bottomrule
   \end{tabular}
\addtolength{\tabcolsep}{-5pt} % set back to normal
\end{table}
%%%%%%%%%%%%%%%%%%%%%%%%%%%%%%%

%bench mark point is encode in BMP-c.tex
%%%%%%%%%%%%%%%%%%%%%%%%%%%%%%%%%%%%%%%%%%%%%%%%%%

%%%%%%%
\subsection{LHC constraints}
In the present setup, direct pair productions of neutralinos, charginos, and sleptons are the targets at LHC searches.
We consider the following channels which have been studied by the ATLAS and CMS collaborations: 
\begin{align}
\text{SLSL:}\quad& pp\to \tilde \ell\w L \tilde \ell\w L^{\ast} \to 
(\ell \neut[1])(\bar \ell \neut[1])\,,
\\
\text{CC/WW:}\quad & pp\to \charP[1]\charM[1]\to
(W^{+}\tilde{\chi}^0_1)(W^{-}\tilde{\chi}^0_1)\,,\label{eq:CC/WW}\\
\text{NC/HW:}\quad & pp\to \neut[2]\charPM[1]\to(h\neut[1])(W^\pm\neut[1])\,,\label{eq:NC/HW}\\
\text{NC/ZW:}\quad & pp\to \neut[2]\charPM[1]\to(Z\neut[1])(W^\pm\neut[1])\,,\label{eq:NC/ZW}\\
\text{NC/3L:}\quad& pp\to \neut[2]\charPM[1]\to 
\begin{cases}
 (l \tilde l\w L)(\nu \tilde{l}\w L)
 \to
 (ll \neut[1])(\nu l \neut[1])\,,
 \\
  (l \tilde{l}\w L) (l \tilde{\nu})
 \to
 (l l \neut[1])(l \nu \neut[1])\,.
\end{cases}
\label{eq:NC/3L}
\end{align}
where $\ell=e,\mu$, $\tilde{\ell}\w L=\tilde{e}\w L,\tilde{\mu}\w L$, $l=e,\mu,\tau$, $\tilde{l}\w L=\tilde{e}\w L,\tilde{\mu}\w L,\tilde{\tau}\w L$, and 
$\tilde{\nu}=\tilde{\nu}_{e,\textrm{L}},\tilde{\nu}_{\mu,\textrm{L}},\tilde{\nu}_{\tau,\textrm{L}}$.
For the details of the analysis procedure, we refer the readers to our previous study~\cite{Endo:2020mqz}.

In the previous work \cite{Endo:2020mqz}, the parameter spaces were constrained by the following analyses:
\begin{itemize}
 \item Ref.~\cite{Aad:2019vnb}      by ATLAS collaboration (SLSL, $139\ifb$),            %1908.08215
 \item Ref.~\cite{Aad:2019vvf}      by ATLAS collaboration (CC/WW and NC/HW, $139\ifb$), %1909.09226
 \item Ref.~\cite{Sirunyan:2017lae} by CMS   collaboration (NC/3L, $35.9\ifb$),          %1709.05406
 \item Ref.~\cite{Aaboud:2018jiw}   by ATLAS collaboration (NC/3L, $36.1\ifb$).          %1803.02762
\end{itemize}
In addition, we found that the newly appeared result,\footnote{%
  We also examined the NC/HW analysis in Ref.~\cite{Aad:2020qnn} and the NC/ZW analysis in Ref.~\cite{Sirunyan:2020eab} based on the full Run~2 data, but no constraints are obtained on our parameter space.
  Other fifteen publications from the ATLAS and CMS collaborations are also taken into account; see \cite{Endo:2020mqz} for the full list of references.
}
\begin{itemize}
 \item Ref.~\cite{Sirunyan:2020eab}    by CMS collaboration (SLSL, $139\ifb$)            %2012.08600
\end{itemize}
is also responsible for the exclusion of our parameter spaces;
it excludes the blue-filled region with labels ``SLSL/C'' in Fig.~\ref{fig:chargino}.
The ``SLSL/A'' blue-filled region has been excluded by the ATLAS counterpart.
The red-filled regions are excluded by the CC/WW and NC/HW channels.
For the NC/3L channel, both of the ATLAS and CMS collaborations assume specific mass spectra of electroweakinos and sleptons, which determine the lepton energies. In terms of a mass difference ratio,
\begin{equation}
 x=\frac{m_{\smuL} - m_{\neut[1]}}{m_{\charPM[1]} - m_{\neut[1]}}\,,
 \label{eq:x}
\end{equation}
the CMS collaboration considers three different mass spectra, $x = (0.05,\,0.5,\,0.95)$, while the ATLAS studies $x=0.5$. In Fig.~\ref{fig:chargino}, the corresponding model points are displayed by the dashed black lines, and the  magenta bars show the regions excluded by the NC/3L channel. Although we have not  investigated the NC/3L bounds for arbitrary $x$ value, it is expected that the bounds on $x=0.05,\,0.5,\,0.95$ are continuously connected with a peak around $x=0.5$ (cf. Ref.~\cite{Endo:2013bba}).

Several preliminary results on conference papers~\cite{ATLAS-CONF-2020-015,CMS-PAS-SUS-19-012,CMS-PAS-SUS-20-003} are not included because numerical data have not been made public.
According to our estimation, they provide additional constraints based on the NC/3L and NC/HW signatures.
The new NC/3L result~\cite{CMS-PAS-SUS-19-012} will confirm that our parameter spaces with $m_{\tilde{\mu}\w L} < m_{\tilde{\chi}^{\pm}_1}$ are strongly disfavored, while the new NC/HW result~\cite{CMS-PAS-SUS-20-003} will exclude more region with $m_{\tilde{\mu}\w L} > m_{\tilde{\chi}^{\pm}_1}$, where the impact depends on the LSP mass.

%%%%%%%%%%%%%%%%%%%%%%%%%%%%%%%%%%%%%%%%%%%%%%%%%%
\section{Bino contributions}
\label{sec:bino}
%%%%%%%%%%%%%%%%%%%%%%%%%%%%%%%%%%%%%%%%%%%%%%%%%%

In the previous section, the wino was assumed to be light to enhance the chargino contributions to the muon $g-2$. 
However, such an assumption is not always necessary for the SUSY contributions to be sizable. 
In this section, we consider another scenario in which the neutralino contribution is dominant. 
In particular, as shown in Eq.~\eqref{eq:BLR}, the bino contribution $\amu[BLR]$ 
becomes large when $\mu\tan\beta$ is large and the bino-like neutralino, the left- and right-handed sleptons are light.

%%%%%
\subsection{Setup}
Let us summarize our setup in this section.
\begin{itemize}
\item The bino-like neutralino $\neut[1]$, left-handed sleptons $\tilde l\w L$, $\tilde\nu$ and right-handed sleptons $\tilde l\w R$ are assumed to be light.
In addition, the higgsinos may be relatively light when $\tan\beta$ is large.
\item The wino is assumed to be decoupled.
Then, the chargino contributions $\amu[WHL]$ discussed in Sec.~\ref{sec:chargino} are suppressed even when $\mu$ is relatively small. 
\item As in the previous section, all the colored SUSY particles and the heavy Higgs bosons are decoupled. 
\item We assume the scalar trilinear terms $(A_e)_{ij}$ to be zero for simplicity.\footnote{The Bino contribution $\amu[BLR]$ can be enhanced by extremely large $(A_e)_{22}$ instead of enlarging $\mu$ and $\tan\beta$.}
\end{itemize}
Then, the following five model parameters are left relevant:
\begin{equation}
    M_1, \quad \mu, \quad (m^2\w L)_i,  \quad (m^2\w R)_i,  \quad  \tan\beta,
    \label{eq:BinoParameters}
\end{equation}
where $m\w L$ and $m\w R$ represent the soft masses for the left- and right-handed sleptons, respectively, with the flavor index $i=\sel,\smu,\stau$.
More concretely, we consider the following two subspace of the parameters:
\begin{itemize}
    \item Universal slepton mass: 
    $(m\w L)_{\sel} = (m\w L)_{\smu} = (m\w L)_{\stau}$ and $(m\w R)_{\sel} = (m\w R)_{\smu} = (m\w R)_{\stau}$.
    \item Heavy stau: 
    $(m\w L)_{\tilde \tau} > (m\w L)_{\tilde \ell}$ and $(m\w R)_{\tilde \tau} > (m\w R)_{\tilde \ell}$.
\end{itemize}
Although $\amu[BLR]$ is independent of the stau mass, some of the constraints discussed below depend on it. 
It will be shown that much wider parameter space is allowed for the heavy stau scenario.

In addition to the one-loop contribution $\amu[BLR]$, there are non-negligible two-loop corrections.
Among them, QED corrections \cite{vonWeitershausen:2010zr} and those to the Yukawa couplings \cite{Marchetti:2008hw,Girrbach:2009uy} are taken into account
in this section.
The latter corrections are also included in calculating the slepton mass spectra and mixing matrices.
In order to treat them consistently when the corrections are large, we do not use \package{GM2Calc} in the following analysis. 
Meanwhile, we neglect radiative corrections to the neutralino couplings, which are not decoupled by heavy SUSY particles (see Refs.~\cite{Nojiri:1996fp,Nojiri:1997ma,Cheng:1997sq,Cheng:1997vy,Katz:1998br,Fargnoli:2013zda,Endo:2013lva}). 
Although the Bino coupling is equal to the gauge coupling at the tree level, this equality is violated by the SUSY-breaking effects, and the deviations are enlarged when the soft SUSY-breaking scale is large.
According to the renormalization group approach~\cite{Endo:2013lva}, they can amount to $5\text{--}10$\% corrections to
$\amu[SUSY]$
when the soft masses, especially the wino mass, are as large as $10\text{--}100\TeV$.
Other higher-order contributions are not-yet-estimated or expected to be smaller.

%%%
\subsection{Constraints}

When the bino and sleptons are light, there are several phenomena which are potentially correlated with the muon $g-2$. 
We consider the following four constraints:
\begin{itemize}
    \item LHC (and LEP) searches for the SUSY particles,
    \item Higgs coupling measurements,
    \item neutralino LSP as a candidate of the dark matter,
    \item stability of the electroweak vacuum.
\end{itemize}
Let us discuss them in turn.
First, similarly to the previous section, the LHC experiment has studied signatures of the direct slepton productions,
\begin{align}
pp
\to \tilde \ell \tilde \ell^{\ast} 
\to (\ell \neut[1])(\bar \ell \neut[1])\,.
\end{align}
From this SLSL channel, where events with energetic two electrons or muons are studied, 
lower bounds on the lightest neutralino mass are obtained as a function of the slepton mass by the ATLAS collaboration ($8\TeV$, $20.3\ifb$) \cite{Aad:2014vma},  %1403.5294
($13\TeV$, $139\ifb$) \cite{Aad:2019vnb}, %1908.08215
and the CMS collaboration ($13\TeV$, $139\ifb$) \cite{Sirunyan:2020eab}.  %2012.08600
Furthermore, 
the ATLAS collaboration has studied the compressed mass spectrum of the neutralino and sleptons.
From the SLSL-soft channel, where events with low-transverse momentum electrons or muons are studied, 
bounds on the slepton masses are obtained for the mass difference $\lesssim 30\GeV$ ($13\TeV$, $139\ifb$) \cite{Aad:2019qnd}.
Furthermore, we impose the LEP bound on the stau mass, $m_{\stau_1} > 95.7$~GeV~\cite{LEP2SUSYWG:04-01.1}.

In addition, events with two energetic hadronically-decaying taus have been studied to search for the direct stau pair productions. 
Although we investigated the latest results by the ATLAS collaboration ($13\TeV$, $139\ifb$) \cite{Aad:2019byo} and the CMS collaboration ($13\TeV$, $77.2\ifb$) \cite{CMS:2019eln}, their constraints were found to be weaker than the above. Also, the smuon/selectron masses are limited by the LEP results \cite{LEP2SUSYWG:04-01.1,LEP2SUSYWG:01-03.1}.
The constraints are weaker than $100\GeV$ and out of the range of our plots. 

The decays of the SM-like Higgs boson have been measured well at the LHC experiment. 
Among the various decay channels, the staus may affect the branching ratio of $h \to \gamma\gamma$ especially when $\mu\tan\beta$ is large and the staus are light
\cite{Carena:2011aa,Carena:2012gp,Kitahara:2012pb,Carena:2012mw}.\footnote{Effects of smuons or selectrons are much weaker because the Yukawa couplings are tiny.}
Currently, the signal strength\footnote{To be exact, 
the production cross section and the total width of the Higgs boson may also be modified by the SUSY particles. 
However, the staus barely affect it, and we ignore such contributions.} has been measured to be \cite{Zyla:2020zbs}
\begin{align}
\mu_{\gamma\gamma} |_{\rm PDG~average}
= \frac{\Gamma(h \to \gamma\gamma)}{\Gamma(h \to \gamma\gamma)_{\rm SM}} 
= 1.11^{+0.10}_{-0.09}\,,
\label{eq:HiggsGammaGamma}
\end{align}
which is consistent with the SM prediction $(\mu^{\rm SM}_{\gamma\gamma}=1)$.\footnote{The CMS collaboration recently reported a preliminary result, $\mu_{\gamma\gamma} =  1.12 \pm 0.09$, using the LHC Run 2 full data. Although it has not been accounted in the above PDG average, the result is consistent with Eq.~\eqref{eq:HiggsGammaGamma}, and the following results are almost unchanged.}
Meanwhile, the measurement is expected to be improved in future; the precision may reach $\delta\mu^{\gamma\gamma}/\mu^{\gamma\gamma} = 3\text{--}4\%$ at HL-LHC ($14\TeV$, $6\iab$) or ILC ($1\TeV$, $8\iab$), 
$2\%$ if they are combined, and would be $0.6\%$ at FCC-ee/eh/hh~\cite{deBlas:2019rxi}.

In the analysis, we require that the lightest neutralino is the LSP.\footnote{The lightest charged slepton is likely to be lighter than sneutrinos when $\mu\tan\beta$ is large enough.}
Although it is one of the best-known candidates of the dark matter,
if it is almost composed of the bino, 
its thermal relic abundance easily exceeds the measured value~\cite{Aghanim:2018eyx,Zyla:2020zbs},
\begin{align}
\Omega_{\rm DM} h^2 = 0.1200 \pm 0.0012\,.
\label{eq:oh2}
\end{align}
In the present setup, the neutralino relic abundance can be consistent with this result by the slepton coannihilation, \ie, when the slepton mass is close to the lightest neutralino appropriately.
We estimate the relic abundance of the LSP neutralino by using the public package \package[5.2.7.a]{micrOMEGAs}~\cite{Belanger:2013oya,Belanger:2010pz,Belanger:2008sj,Belanger:2006is}.

Such a dark matter  may be detected by the direct detection. 
The LUX~\cite{Akerib:2016vxi}, PandaX-II~\cite{Cui:2017nnn} and XENON1T~\cite{Aprile:2018dbl} experiments have provided stringent constraints on the spin-independent cross section of the dark matter scattering off the nuclei. 
In the analysis, the cross section is estimated by using \package{micrOMEGAs}.

The last constraint is the vacuum meta-stability condition.
The trilinear coupling of the sleptons and the SM-like Higgs boson is given by
\begin{align}
V & \simeq 
-\frac{m_{\ell}}{\sqrt{2}v(1+\Delta_\ell)}\mu\tan\beta 
\cdot\tilde{\ell}_{L}^* \tilde{\ell}_{R} h + {\rm h.c.},
\end{align}
where $v \simeq 174\GeV$ is the vacuum expectation value of the SM-like Higgs, and $\Delta_\ell$ represents the radiative corrections to the lepton Yukawa couplings mentioned above~\cite{Girrbach:2009uy}. 
As $\mu\tan\beta$ increases, the trilinear coupling is enhanced, charge-breaking minima become deeper, and eventually, the stability of the electroweak vacuum is spoiled. 
We require the lifetime of the vacuum
being longer than the age of the Universe, restricting $|\mu\tan\beta|$ from above. For the following analysis, 
we adopt the fitting formula explored in Refs.~\cite{Kitahara:2013lfa,Endo:2013lva}, 
\begin{equation}
\begin{split}
&\left|
    \frac{m_{\ell}\,\mu\tan\beta }{\sqrt{2}v(1+\Delta_\ell)}
    \right|
    \\&\quad
\le
    \eta_\ell
    \bigg\{
 1.01  \times 10^2 \GeV \sqrt{(m\w L)_{\tilde\ell} (m\w R)_{\tilde\ell}} 
 + 1.01 \times 10^2 \GeV  \left[ (m\w L)_{\tilde\ell}  + 1.03 (m\w R)_{\tilde\ell} \right] 
  \\&\qquad
 + \frac{2.97 \times 10^6\GeV^3}{(m\w L)_{\tilde\ell}  + (m\w R)_{\tilde\ell} } 
 - 1.14 \times 10^8\GeV^4 
  \left[ \frac{1}{(m^2\w L)_{\tilde\ell} } +  \frac{0.983}{(m^2\w R)_{\tilde\ell}} \right] 
   -2.27 \times 
 10^4\GeV^2
 \bigg\}\,,
\end{split}    
\label{eq:vacuum_condition}
\end{equation}
with $\ell = e, \mu, \tau$.
The numerical evaluation was done by the bounce method~\cite{Coleman:1977py} via \package[1.0.2]{CosmoTransitions}~\cite{Wainwright:2011kj}.\footnote{Note that we ignore thermal effects and radiative corrections to the formula, which may affect the following result~\cite{Endo:2010ya,Endo:2015ixx} and will be discussed elsewhere. }
A factor $\eta_\ell \sim 1$ 
provides a mild $\tan \beta$ dependence coming from the 
Yukawa contributions to the quartic terms in the scalar potential.
The explicit value of $\eta_\ell$ is found in Refs.~\cite{Kitahara:2013lfa,Endo:2013lva}.

%%%%%%%%%%%%%%%%%%%%%%%%%%%%%%%%%%%%%%%%%%%%%%%%%%
\subsection{Result in universal slepton mass case}
%%%%%%%%%%%%%%%%%%%%%%%%%%%%%%%%%%%%%%%%%%%%%%%%%%

\begin{figure}[p]
 \centering
 \renewcommand\thesubfigure{\Alph{subfigure}}
%%%%%
  \begin{subfigure}[b]{0.49\textwidth}
 \includegraphics[width=\textwidth]{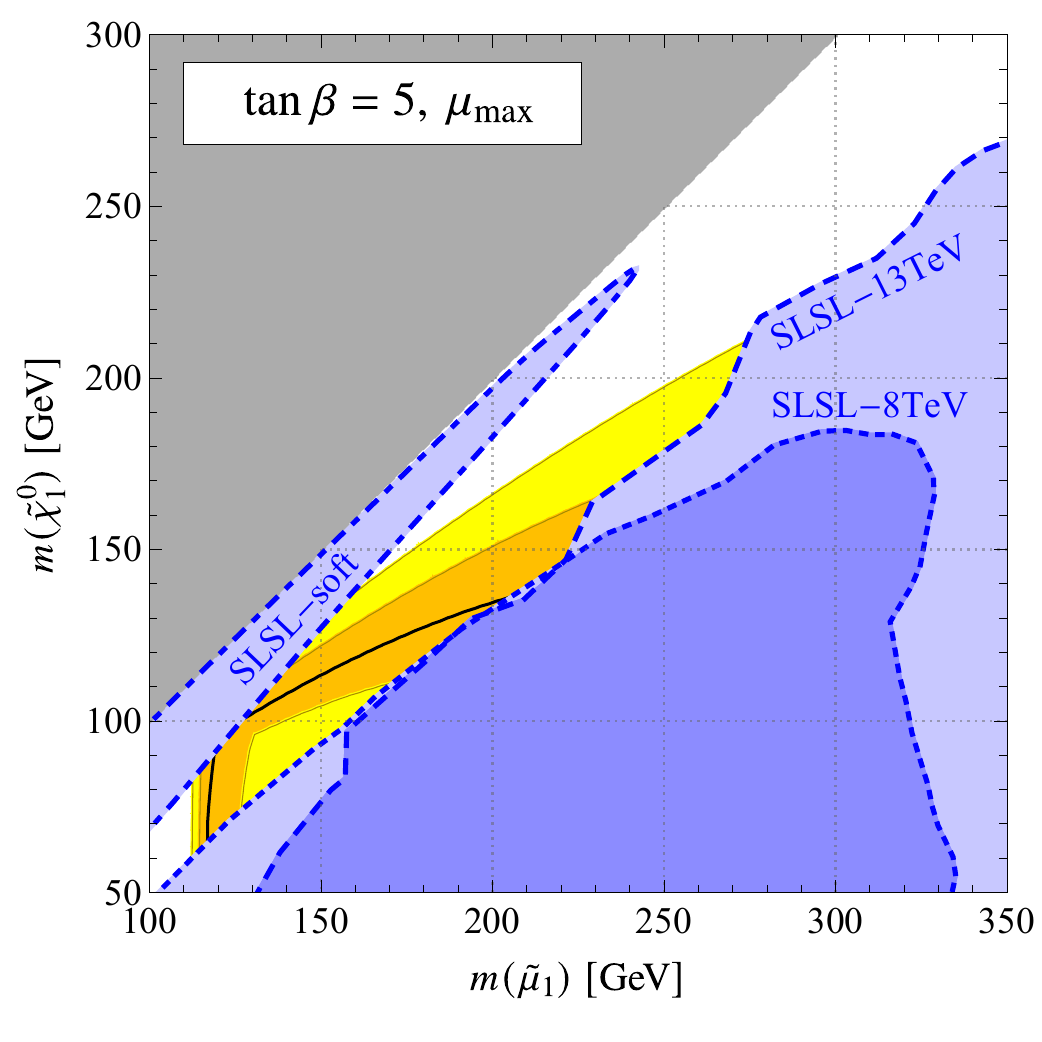}
\caption{$\tan\beta=5$}
 \end{subfigure}
 %%%
  \begin{subfigure}[b]{0.49\textwidth}
 \includegraphics[width=\textwidth]{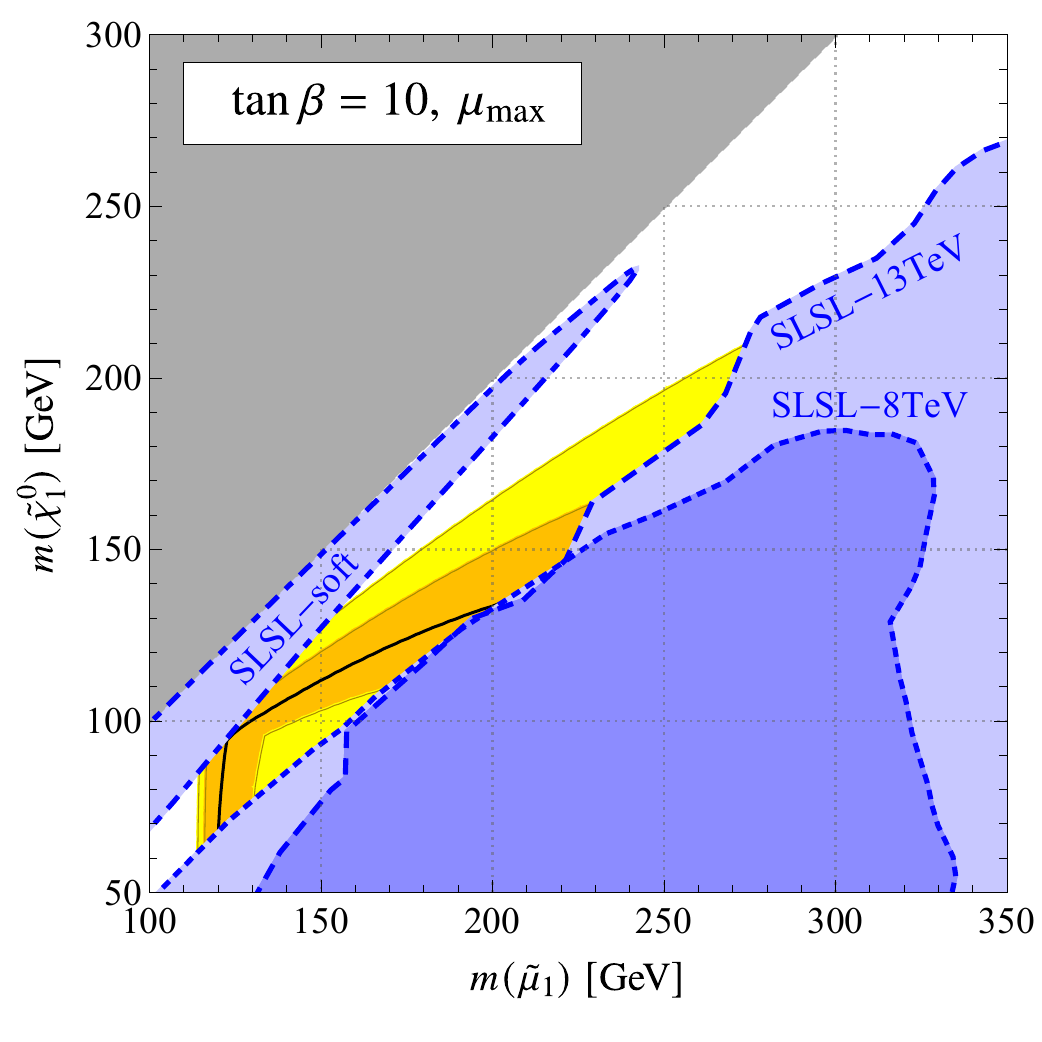}
\caption{$\tan\beta=10$}
 \end{subfigure}
\par\vspace{2em} %%% textheightをexceedsしたらここを減らして
   \begin{subfigure}[b]{0.49\textwidth}
 \includegraphics[width=\textwidth]{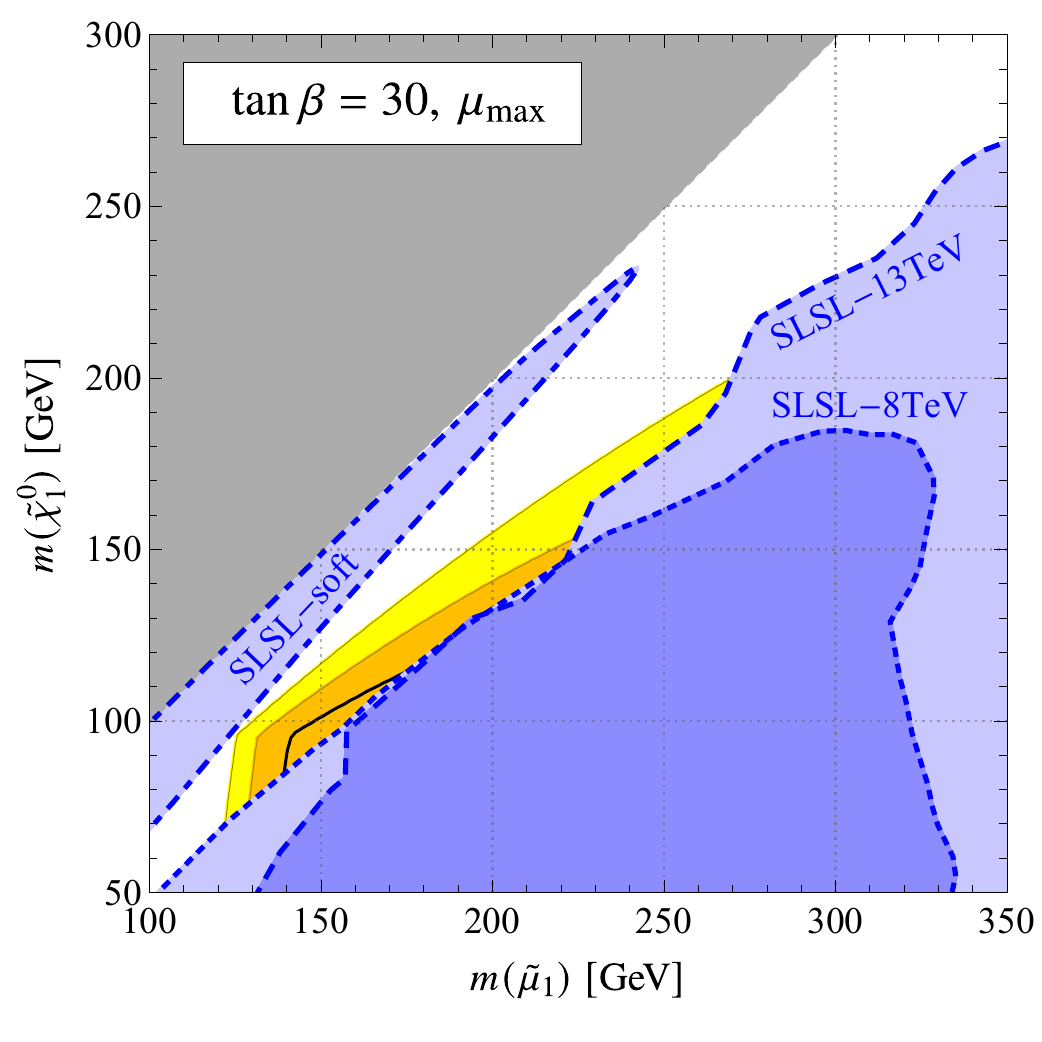}
\caption{$\tan\beta=30$}
 \end{subfigure}
 %%%
   \begin{subfigure}[b]{0.49\textwidth}
 \includegraphics[width=\textwidth]{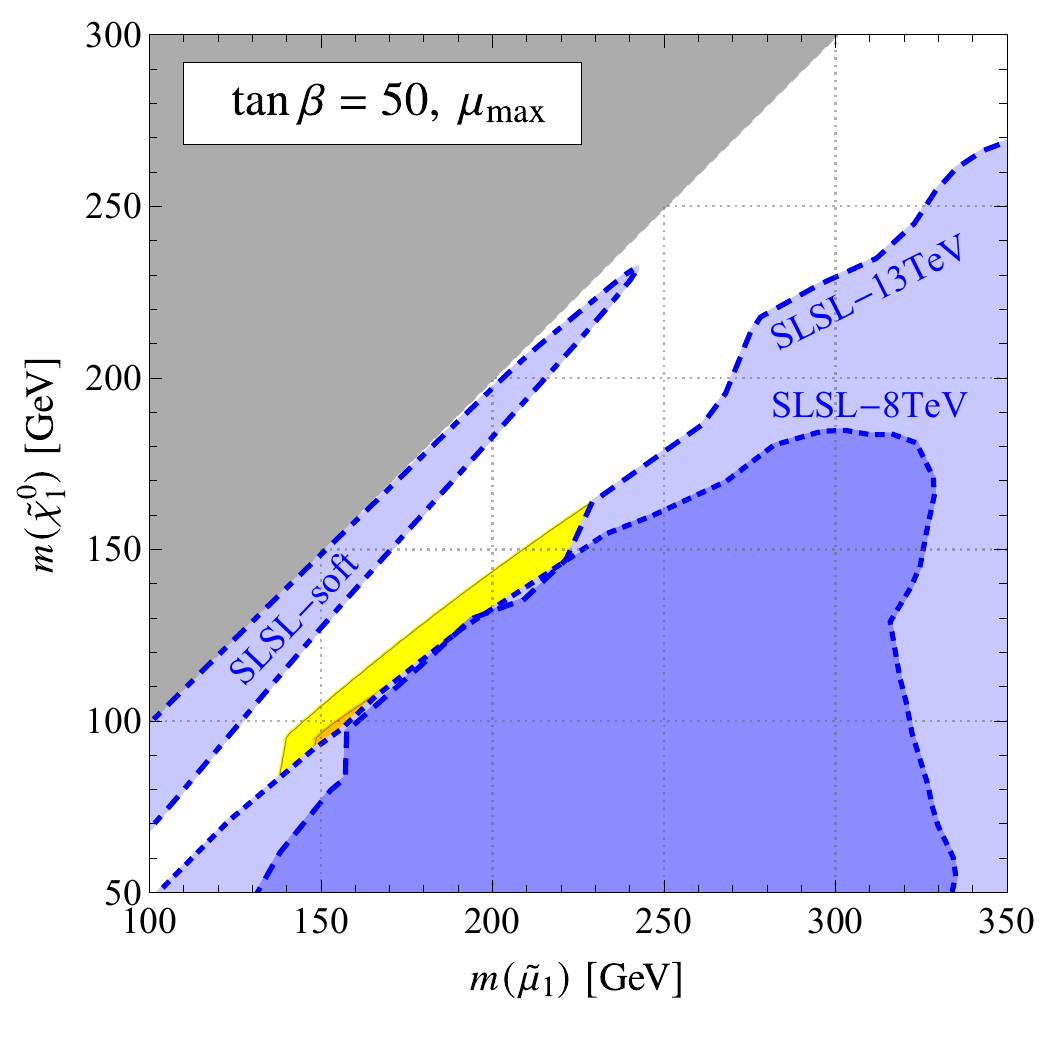}
\caption{$\tan\beta=50$}
 \end{subfigure}
 %%%%%
  \caption{\label{fig:neutralino}%
  The   summary  of  the  bino-dominated SUSY  scenario  for  the muon $g-2$ anomaly. 
  The universal slepton mass with $m\w L = m\w R$ 
  is assumed.
  Four planes respectively correspond to $\tan\beta = 5,10,30$ and $50$.
  The $\mu$ parameter is maximized $(\mu_{\rm max})$ at each point  
  under the conditions described in the text.
  The muon $g-2$ anomaly can be explained at the $1\sigma$ ($2\sigma$) level in the orange-filled (yellow-filled) regions.
    Below the black line in each figure,  $\amu[SUSY]$ exceeds the central value of $\Delta \amu$ in Eq.~\eqref{eq:Deltaamu} for 
    the maximized $\mu$ parameter.
    In the gray-filled regions, $\tilde\mu_1$ is lighter than $\neut[1]$.
    The blue-filled regions are excluded by the LHC slepton searches~\cite{Aad:2014vma, Aad:2019vnb, Sirunyan:2020eab, Aad:2019qnd}.
  }
\end{figure}

Let us first consider the slepton universal mass case. 
The left-handed (also right-handed) selectron, smuon and stau have a common soft SUSY-breaking mass,
\begin{align}
 (m\w L)_{\sel} = (m\w L)_{\smu} = (m\w L)_{\stau}\,, \quad
 (m\w R)_{\sel} = (m\w R)_{\smu} = (m\w R)_{\stau}\,.
\end{align}
Such a spectrum is motivated to avoid dangerous lepton flavor or CP violations (see Ref.~\cite{Endo:2013lva}).
The SUSY contribution to the muon $g-2$ cannot be arbitrarily large because $\mu\tan\beta$ is bounded from above;
due to the stau left-right mixing, 
a too large $\mu\tan\beta$ can violate one (or more) of the following constraints, (i) the vacuum stability in the stau--Higgs potential, Eq.~\eqref{eq:vacuum_condition}, (ii) the neutralino being the LSP, \ie, $m_{\stau_1} > m_{\neut[1]}$, and (iii) the LEP bound on the stau mass, $m_{\stau_1} > 95.7$~GeV.
Thus, these conditions yield an upper bound on the $\mu$ parameter as a function of $M_1$, $(m\w L)_{\stau}$, and $(m\w R)_{\stau}$.

Figure~\ref{fig:neutralino} shows the results for $m\w L = m\w R$ and $\tan\beta = 5, 10, 30, 50$. 
The horizontal and vertical axes are the physical masses of the lightest neutralino $\neut[1]$ and the lighter smuon $\tilde\mu\w 1$, respectively.
For a given set of $m\w L (= m\w R)$, $M_1$, and $\tan\beta$, the $\mu$ parameter (or equivalently $\mu\tan\beta$) is maximized under the above conditions, so that $\amu[BLR]$ becomes maximum. 
In the orange-filled (yellow-filled) regions, the SUSY contribution explains the muon $g-2$ discrepancy at the $1\sigma$ ($2\sigma$) level.
Below the black line in each figure,  $\amu[SUSY]$ exceeds the central value of $\Delta \amu$ [Eq.~\eqref{eq:Deltaamu}] for the maximized $\mu$ parameter.
In other words, in these regions, $\amu[SUSY]$ can be optimized by reducing the $\mu$ parameter.
The upper edges of these muon $g-2$ regions, \ie, the upper bounds on $m_{\neut[1]}$, are 
determined by the condition $m_{\stau_1} > m_{\neut[1]}$.\footnote{To be exact, $\stau_1$ becomes long-lived if $m_{\stau_1}$ is too much close to $m_{\neut[1]}$. Thus, $m_{\stau_1}-m_{\neut[1]} \gtrsim 2\GeV$ is required in practice, though the figures are almost unchanged. } 
In the low mass region of $m_{\neut[1]}\lesssim 100\GeV$, the LEP bound on the stau mass also affects the boundary.
Meanwhile, without the LHC constraints, there would be lower boundaries (\ie, lower bounds on $m_{\neut[1]}$, or equivalently upper bounds on $m_{\smu_1}$) due to the vacuum meta-stability condition, though they are hidden by the LHC constraints from the SLSL channel (lower blue-filled regions).
Here, the region with smaller $m_{\smu_1}$ is constrained by the ATLAS $8\TeV$ result, while the larger one is by the ATLAS $13\TeV$ result.
Besides, the SLSL-soft channel restricts the regions of degenerate neutralino-smuon masses (upper blue-filled region).

In Fig.~\ref{fig:neutralino} (A), (B), which correspond to $\tan\beta = 5, 10$ respectively, $\smu_1$ is required to be $m_{\tilde{\mu}_1} \lesssim 230~(270)\GeV$ 
to explain the muon $g-2$ discrepancy at the $1\sigma$ ($2\sigma$) level. 
It is noticed that the regions become narrower as $\tan\beta$ increases (see Fig.~\ref{fig:neutralino} (C), (D) for $\tan\beta = 30, 50$). 
The reason is as follows. 
At each set of $(M_1, m\w L=m\w R)$, the upper bound on $\mu\tan\beta$ is almost unchanged. 
Then, since larger $\tan\beta$ leads to smaller $\mu$, the BHR contribution to $\amu[SUSY]$, which destructively interferes with $\amu[BLR]$, becomes enhanced when $\mu$ is smaller, \ie, $\tan\beta$ is large. 
Note that the BHL contribution is weaker than that of BHR for $m\w L \sim m\w R$ [see Eqs.~\eqref{eq:BHL} and \eqref{eq:BHR}].
Consequently, lower $\tan\beta$ is favored to enhance $\amu[SUSY]$.
Note that $\mu \sim 500\GeV\text{--}4\TeV$ for $\tan\beta=10$ in the muon $g-2$ region of Fig.~\ref{fig:neutralino} (A), which are large enough to suppress $\amu[BHL]$ and $\amu[BHR]$.

The branching ratio of $h \to \gamma\gamma$ is affected by light staus with large $\mu\tan\beta$. 
The signal strength is likely to be enhanced in the vicinity of the boundary of the SLSL constraint, while it is weaker for larger $m_{\neut[1]}$ as the upper bound on $\mu\tan\beta$ is tighter.\footnote{
Note that contributions to $\mu_{\gamma\gamma}$ coming from light chargino loops may amount to additional few$\,\%$ at most.
Such contributions, however, are scaled by $\approx 1/(\mu M_2 \tan \beta)$~\cite{Batell:2013bka,Endo:2014pja}.
Hence, they are certainly suppressed by $M_2$ in the present setup.}
Around the muon $g-2$ regions in the figures, it can be deviated from the SM value $\mu_{\gamma\gamma}$ by $2$--$5\%$ at most, which satisfies the current limit \eqref{eq:HiggsGammaGamma} and is accessible in future by HL-LHC, ILC, or FCC-ee/eh/hh.

\begin{figure}[t]
 \centering
 \renewcommand\thesubfigure{\Alph{subfigure}}
%%%%%
 %%%
   \begin{subfigure}[b]{0.49\textwidth}
 \includegraphics[width=\textwidth]{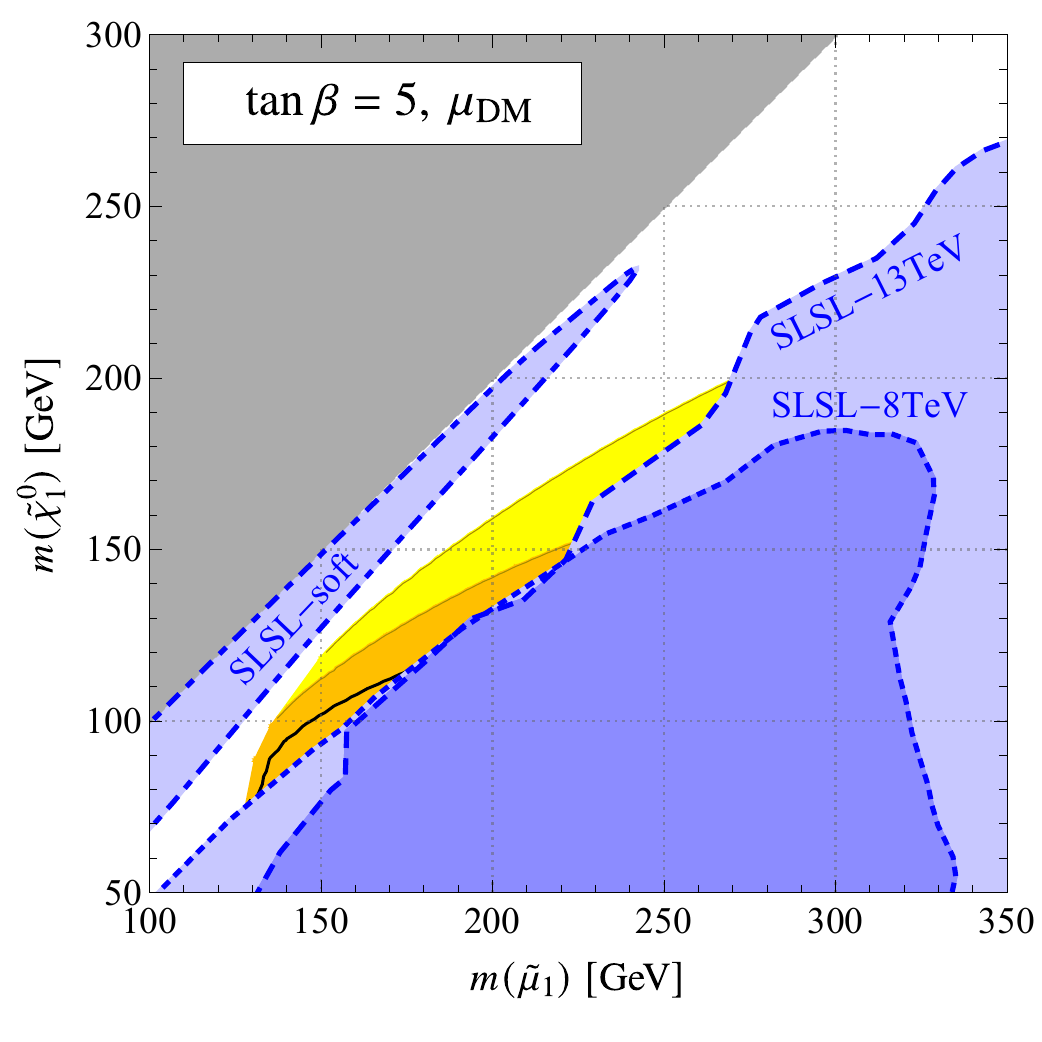}
\caption{$\tan\beta=5$}
 \end{subfigure}
 %%%
   \begin{subfigure}[b]{0.49\textwidth}
  \includegraphics[width=\textwidth]{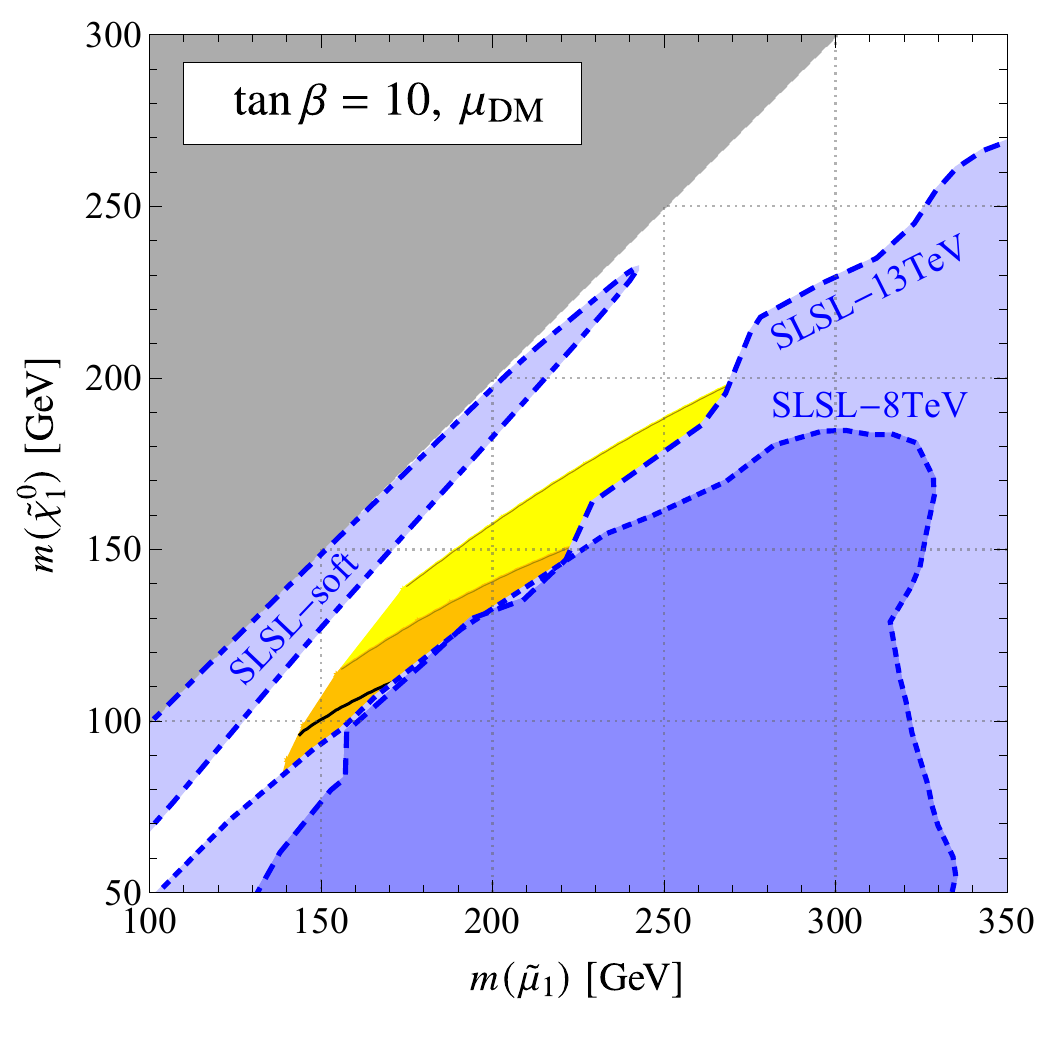}
\caption{$\tan\beta=10$}
 \end{subfigure}
 %%%%%
  \caption{\label{fig:neutralinoDM}%
  The bino-dominated SUSY scenario for the muon $g-2$ anomaly
  with proper dark matter relic abundance.
  In the same way as Fig.~\ref{fig:neutralino}, 
  the universal slepton mass with $m\w L = m\w R$ 
  are used, but the $\mu$ parameter is tuned ($\mu_{\rm DM}$) to realize 
  $\Omega_{\rm DM}h^2=0.12$.
  We show cases of $\tan\beta =5$ and $10$.
In the orange-filled (yellow-filled) regions, the muon $g-2$ anomaly can be solved at the $1\sigma$ ($2\sigma$) level while satisfying the XENON1T constraint.
  }
\end{figure}

In Fig.~\ref{fig:neutralino}, the thermal relic abundance of $\neut[1]$ is not tuned to reproduce the dark matter value \eqref{eq:oh2}. 
After imposing the LHC constraints, we found that the abundances in the muon $g-2$ region satisfy $\Omega_{\neut[1]} < \Omega_{\rm DM}$ because the stau masses are well close to that of $\neut[1]$.
Let us next consider the case where $\mu\tan\beta$ is determined by requiring that the thermal relic abundance of the neutralino is equal to Eq.~\eqref{eq:oh2}.
It is almost determined by the coannihilations with the lightest stau.\footnote{There can be more than one values of the $\mu$ parameter that realize the desired dark matter density; in such a case, we take the largest $\mu$ 
 as it gives the largest contribution to the muon $g-2$.}
In Fig.~\ref{fig:neutralinoDM}, the muon $g-2$ discrepancy is explained at the $1\sigma$ ($2\sigma$) level in the orange-filled (yellow-filled) regions 
while satisfying $\Omega_{\neut[1]}h^2=0.12$.
Here, $\tan\beta = 5$ (A) and $10$ (B) with $m\w L = m\w R$.
It is found that $\smu_1$ is required to be $m_{\tilde{\mu}_1} \lesssim 220~(270)\GeV$ 
to explain the muon $g-2$ discrepancy at the $1\sigma$ ($2\sigma$) level.

In the figures, the left boundaries of the orange-filled (yellow-filled) regions for small $m_{\neut[1]}$ are determined by the XENON1T constraint of the dark matter direct detection~\cite{Aprile:2018dbl}. 
Here, $\mu$ cannot be so large, and thus, 
the dark matter-nucleon elastic scattering cross-section is enhanced.
For smaller $\tan\beta$, wider parameter regions are allowed because the coannihilation works efficiently by larger $\mu$, \ie, without the well-tempered contributions. 
In the near future, the sensitivity of the direct detection is planned to be improved by orders of magnitudes.
We have checked that most of the muon $g-2$ parameter regions can be probed by the XENONnT experiment~\cite{Aprile:2020vtw} if the slepton mass spectrum is universal.

We also found that the branching ratio of the SM-like Higgs boson decay to two photons, $\mu_{\gamma\gamma}$, can deviate from the SM prediction by at most 2\% in the muon $g-2$ parameter region.

In Table~\ref{tab:BMP}, we show the mass spectra, the SUSY contribution to the muon $g-2$ ($\amu[SUSY]$), the thermal relic abundance of $\neut[1]$ ($\Omega_{\neut[1]}h^2$), the spin-independent cross section of the neutralino for the dark matter direct detection ($\sigma_p^{\rm SI}$), and the branching ratio of $h \to \gamma\gamma$ ($\mu_{\gamma\gamma}$)
for some benchmark points in Fig.~\ref{fig:neutralinoDM}.
The points BLR1 and BLR2 can be probed by the direct stau searches at the ILC $250\GeV$, 
while all the four points are accessible by the ILC $500\GeV$ as well as the XENONnT experiment.
Furthermore, all the points could be tested by measuring the branching ratio of the SM-like Higgs boson decay at FCC-ee/eh/hh.

%%%%%%%%%%%%%%%%%%%%%%%%%%%%%%%%%%%%%%%%%%%%%%%%%%
%%%%%%%%%%%%%%%%%%%%%%%%%%%%%%%
\begin{table}[t]
\centering
\newcommand{\bhline}[1]{\noalign{\hrule height #1}}
\renewcommand{\arraystretch}{1.5}
\rowcolors{2}{gray!15}{white}
\addtolength{\tabcolsep}{5pt} % add space between columns
\caption{\label{tab:BMP} 
Benchmark points for the pure-bino-contribution dominated scenario, where the universal slepton mass is used.
The mass parameters are in units of GeV.
}
  \begin{tabular}{c cccc} 
  %\toprule
  \bhline{1 pt}
  \rowcolor{white}
  & BLR1 & BLR2 & BLR3 & BLR4 \\  \hline 
 $M_1$ & 100 & 100 & 150 & 150  \\  
 $m\w L = m\w R$ & 150 & 150 & 200 & 200 \\ 
 $\tan \beta$ & 5 & 10 & 5 & 10 \\   
 $\mu$ & 1323 & 678 & 1922 & 973 \\   \hline 
 $m_{\tilde{\mu}_1} $ & 154 & 154 & 202 & 202 \\ 
 $m_{\tilde{\mu}_2} $ & 159 & 159 & 207 & 208 \\ 
 $m_{\tilde{\tau}_1}$ & 113 & 113 & 159 & 158 \\ 
 $m_{\tilde{\tau}_2}$ & 190 & 191 & 242 & 243 \\ 
 $m_{\tilde{\nu}_{\mu,\tau}}$ & 137 & 136 & 190 & 190 \\ 
 $m_{\tilde{\chi}_1^0}$ & 99 & 99 & 150 & 149 \\ 
 $m_{\tilde{\chi}_2^0}, m_{\tilde{\chi}_3^0}, m_{\tilde{\chi}_1^{\pm}}$ & 1323--1324 & 678--680 & 1922--1923 & 973--975 \\
 \hline 
 $\amu[SUSY]\times 10^{10}$ & 27 & 27 & 17 & 17 \\ 
 $\Omega_{\rm DM} h^2$ & 0.120 & 0.120 & 0.120 & 0.120 \\ 
 $\sigma_{p}^{\rm SI} \times 10^{47}$ $[{\rm cm}^2] $ & 1.7 & 3.7 & 0.8 & 1.9 \\ 
 $\mu_{\gamma\gamma} $ & 1.01 & 1.01 & 1.01 & 1.01 \\ 
\bhline{1 pt}
%\bottomrule
   \end{tabular}
\addtolength{\tabcolsep}{-5pt} % set back to normal
\end{table}
%%%%%%%%%%%%%%%%%%%%%%%%%%%%%%%

%bench mark point is encode in BMP.tex
%%%%%%%%%%%%%%%%%%%%%%%%%%%%%%%%%%%%%%%%%%%%%%%%%%

So far, we have assumed that the soft masses of the left- and right-handed sleptons have a common value, $m\w L = m\w R$.
According to the loop function of $\amu[BLR]$, for a fixed value of $\smu_1$ the SUSY contributions to the muon $g-2$ is maximized when $m\w L = m\w R$, \ie, the muon $g-2$ regions in Fig.~\ref{fig:neutralino}~\cite{Endo:2013lva}.
Meanwhile, the stau coannihilation is realized by larger $\mu$ for $m\w L \neq m\w R$, which enhances $\amu[BLR]$ simultaneously.
Consequently, the muon $g-2$ regions in Fig.~\ref{fig:neutralinoDM} are enlarged for $m\w L \neq m\w R$, though we need detailed studies for the decay of the heavier sleptons to the lightest neutralino, which may be excluded by the SLSL search at the LHC.

%%%%%%%%%%%%%%%%%%%%%%%%%%%%%%%%%%%%%%%%%%%%%%%%%%
\subsection{Result in heavy stau case}
%%%%%%%%%%%%%%%%%%%%%%%%%%%%%%%%%%%%%%%%%%%%%%%%%%

\begin{figure}[t]
 \centering
 \renewcommand\thesubfigure{\Alph{subfigure}}
%%%%%
  \begin{subfigure}[b]{0.49\textwidth}
 \includegraphics[width=\textwidth]{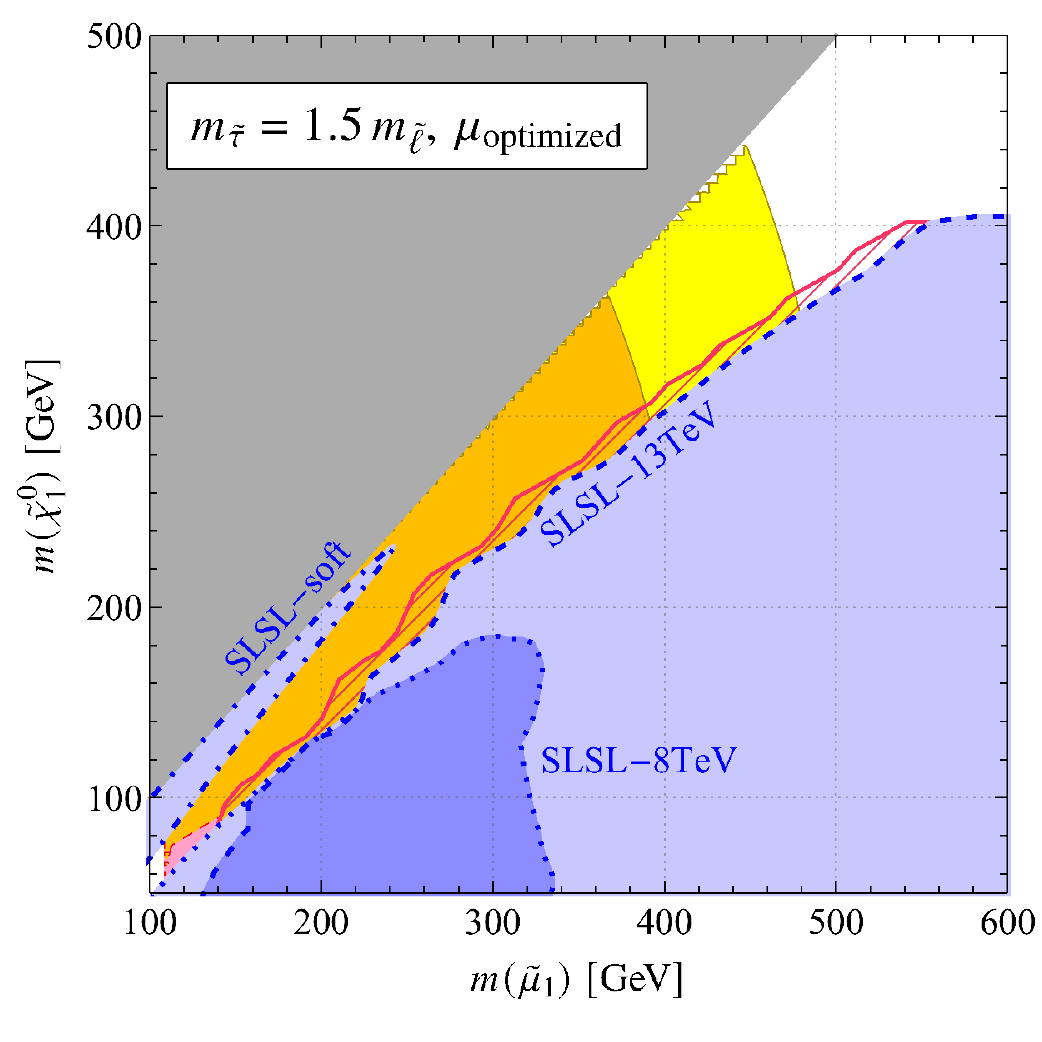}
\caption{$(m\w {L,R})_{\tilde \tau} = 1.5 \,(m\w {L,R})_{\tilde \ell}$}
 \vspace{.2cm}
 \end{subfigure}
 %%%
   \begin{subfigure}[b]{0.49\textwidth}
 \includegraphics[width=\textwidth]{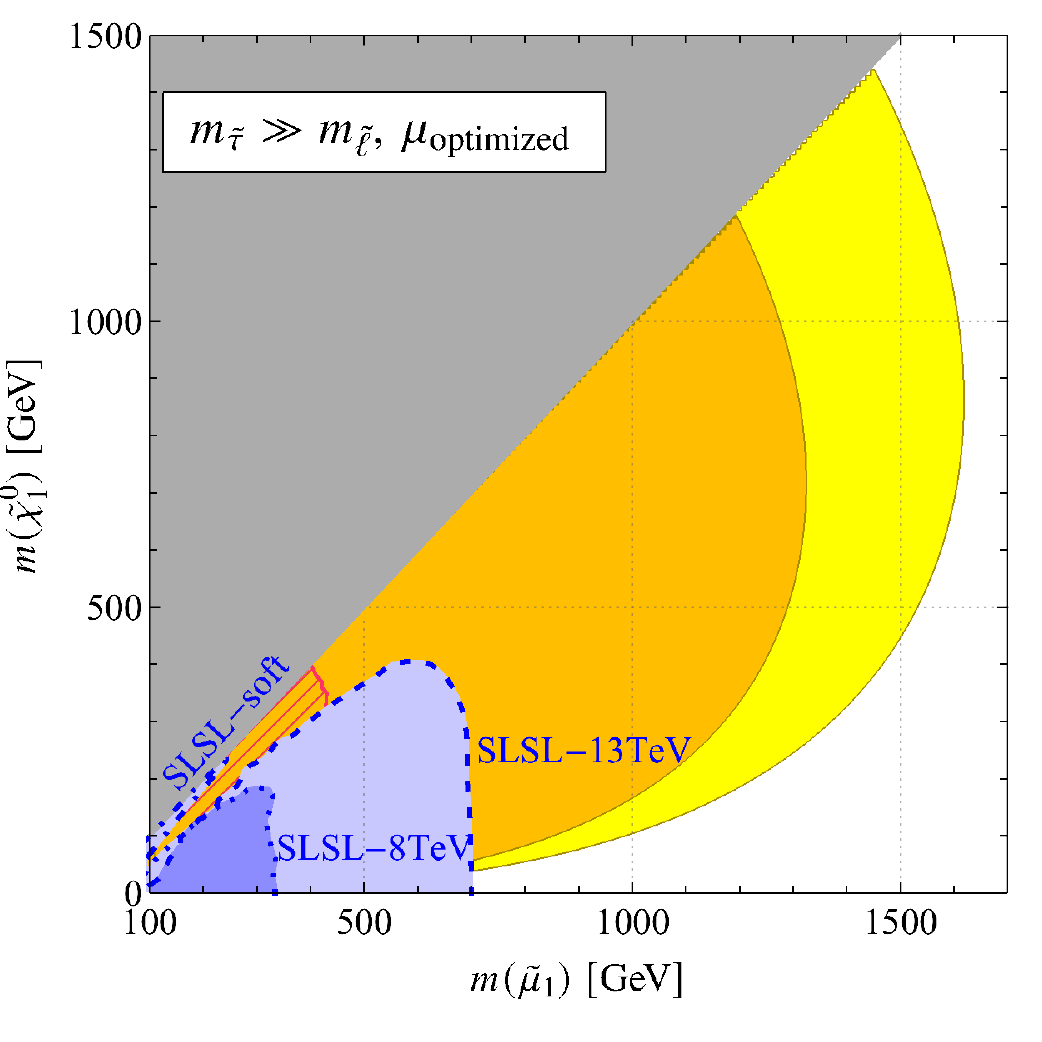}
\caption{$(m\w {L,R})_{\tilde \tau} \gg (m\w {L,R})_{\tilde \ell}$}
 \vspace{.2cm}
 \end{subfigure}
 %%%%%
  \caption{\label{fig:neutralino_heavystau}%
  Same as Fig.~\ref{fig:neutralino}, but the stau soft masses are larger than those of selectrons and smuons ($\tilde\ell = \sel, \smu$). 
  Here, $m\w L = m\w R$ and $\tan\beta=10$.
  The $\mu$ parameter is optimized ($\mu_{\rm optimized}$) at each point to explain the muon $g-2$ discrepancy at the $1\sigma$ ($2\sigma$) level in the orange-filled (yellow-filled) region. 
  The red-hatched region is potentially excluded by the SLSL search for the decay of the heavier smuon to the lightest neutralino (see the text).
  }
\end{figure}

Next, we discuss the case that the staus are heavier than selectrons and smuons,\footnote{It is assumed that dangerous flavor or CP violations are suppressed by some mechanisms.
}
\begin{align}
 (m\w L)_{\sel} = (m\w L)_{\smu} < (m\w L)_{\stau}\,, \quad
 (m\w R)_{\sel} = (m\w R)_{\smu} < (m\w R)_{\stau}\,.
\end{align}
Although the stau masses are irrelevant for the SUSY contribution to the muon $g-2$, the vacuum meta-stability condition of the stau--Higgs potential as well as the condition $m_{\stau_1} > m_{\neut[1]}$ are relaxed as they increase.
For $m_{\stau} \gg m_{\tilde\ell}$ with $\tilde\ell = \sel, \smu$, the vacuum meta-stability constraint from the smuon--Higgs potential becomes severer, and the latter condition is replaced by $m_{\smu_1} > m_{\neut[1]}$.

Figure~\ref{fig:neutralino_heavystau} shows the results for (A) $(m\w {L,R})_{\tilde \tau} = 1.5 (m\w {L,R})_{\tilde \ell}$ and (B) $(m\w {L,R})_{\tilde \tau} \gg  (m\w {L,R})_{\tilde \ell}$.
Here, each slepton satisfies $(m\w L)_i = (m\w R)_i$ with $\tan\beta = 10$. 
In the orange-filled (yellow-filled) region, the SUSY contribution can explain the muon $g-2$ discrepancy at the $1\sigma$ ($2\sigma$) level. 
If $\mu\tan\beta$ is maximized under the above conditions,
$\amu[SUSY]$ becomes too large compared with the observed discrepancy in most of the orange-filled region. 
Therefore, $\mu$ is understood to be optimized at each point of these figures. 
The upper boundaries of the regions (the upper bound on $m_{\neut[1]}$) is determined by $m_{\smu_1} > m_{\neut[1]}$, and the right boundaries (the upper bound on $m_{\smu_1}$) is given by the vacuum meta-stability condition of the stau--Higgs potential in (A) and the smuon--Higgs potential in (B).
As a result, it is found that $\smu_1$ is required to be $m_{\tilde{\mu}_1} \lesssim 390~(490)\GeV$ 
to explain the muon $g-2$ discrepancy at the $1\sigma$ ($2\sigma$) level for $(m\w {L,R})_{\tilde \tau} = 1.5 (m\w {L,R})_{\tilde \ell}$.
As the stau becomes heavier, these values become larger.
In the limit of $(m\w {L,R})_{\tilde \tau} \gg (m\w {L,R})_{\tilde \ell}$, one obtains $m_{\tilde{\mu}_1} \lesssim 1.3~(1.6)\TeV$.

The blue-filled regions are excluded by the SLSL and SLSL-soft channels on the lighter smuon, and the magenta-filled region is excluded by $h \to \gamma\gamma$ [Eq.~\eqref{eq:HiggsGammaGamma}] in Fig.~\ref{fig:neutralino_heavystau} (A).
In addition, the heavier smuon is potentially excluded by the SLSL search.
As the staus become heavier than the smuons, $\mu\tan\beta$ is allowed to become larger, and thus, a mass splitting between the lighter and heavier smuons is enlarged especially when $m\w L = m\w R$.
In the red-hatched region, the decay of the heavier smuon to the lightest neutralino is excluded if we suppose that ${\rm BR}(\smu_2 \to \mu \neut[1])=1$ and that $\mu\tan\beta$ is maximized under the above conditions.\footnote{In the universal slepton mass case, the smuon mass splitting is small enough. }
Although a smaller value is enough to explain the muon $g-2$ discrepancy, detailed LHC analysis is necessary to check the viability of this region.

In Fig.~\ref{fig:neutralino_heavystau}, the relic abundance of $\neut[1]$ is not tuned to reproduce the dark matter value \eqref{eq:oh2}, but can exceed it.
In order to obtain $\Omega_{\neut[1]} = \Omega_{\rm DM}$ the smuon is required to be degenerate with the LSP appropriately for the annihilation to work. 
We have checked that the dark matter abundance can be reproduced for $m_{\smu_1} - m_{\neut[1]} \lesssim 0.1 m_{\neut[1]}$ 
while keeping the explanation to
the muon $g-2$ discrepancy, where the coannihilation works with the charged smuons, selectrons, and/or sneutrinos. 
We also found that the dark matter-nucleon elastic scattering cross-sections are much smaller than the XENONnT sensitivity in the heavy stau scenario.

%%%%%%%%%%%%%%%%%%%%%%%%%%%%%%%%%%%%%%%%%%%%%%%%%%
\section{Conclusions and Discussion}
\label{sec:conclusion}
%%%%%%%%%%%%%%%%%%%%%%%%%%%%%%%%%%%%%%%%%%%%%%%%%%

The first result of the measurement of the muon anomalous magnetic moment ($g-2$) by the Fermilab Muon $g-2$ collaboration confirmed the previous result at the Brookhaven National Laboratory
and thus the long-standing discrepancy with the Standard Model prediction.
In this paper, we revisited low-scale supersymmetric models as a solution to this anomaly, focusing on two distinct scenarios; one in which the chargino contribution, $\amu[WHL]$, is dominant, and the other in which the pure-bino contribution, $\amu[BLR]$, is dominant.

In the chargino-contribution dominated scenario, we revisited our previous study~\cite{Endo:2020mqz}, taking account of the latest LHC results as well as the new $\amu$ measurements~\cite{Abi:2021gix} and the new theory combination~\cite{Aoyama:2020ynm}. 
It was found that models with $m_{\tilde{\mu}\w L} < m_{\tilde{\chi}^{\pm}_1}$ in this scenario are disfavored as a solution to the muon $g-2$ anomaly, while models with $m_{\tilde{\mu}\w L} > m_{\tilde{\chi}^{\pm}_1}$ are still widely allowed.
Several benchmark points are listed in Table~\ref{tab:BMP-c}, for which $\amu[SUSY]$ is sizable and the LHC constraints are still allowed.

In the pure-bino-contribution dominated scenario, we first studied the universal slepton-mass case where 
the $\mu$ parameter is maximized under the constraints of the vacuum stability, the neutralino being the LSP, and the LEP bound on the stau mass. 
Although the LHC slepton searches are tight, there remain viable parameter spaces for $m_{\neut[1]} \lesssim m_{\smu_1} \lesssim 300\GeV$: The muon $g-2$ anomaly can be explained within 1$\sigma$ (2$\sigma$) for $m_{\smu_1}\lesssim 230~(270)\GeV$.
Interestingly, the region with low $\tan \beta$ with heavy higgsinos is preferred, because the destructive BHR contribution is suppressed. 

Furthermore, the thermal relic of the bino-like neutralino can become the dominant component of the dark matter, if the stau--bino coannihilation works properly. 
It was shown that the relic abundance as well as the muon $g-2$ anomaly can be explained simultaneously for $m_{\smu_1}\lesssim 220~(270)\GeV$.
Besides, such parameter regions satisfy the XENON1T constraint on the dark matter scattering cross section. 
We list several benchmark points in Table~\ref{tab:BMP}, in which the SUSY contribution to the muon $g-2$ is sizable, the neutralino relic abundance is consistent with the measured dark matter value, and all the constraints are satisfied.

We also investigated the setup that the staus are (much) heavier than the selectrons and smuons. 
Because the conditions of the vacuum meta-stability in the scalar potential 
and that the sleptons are heavier than the lightest neutralino are alleviated drastically, the parameter regions that explain the muon $g-2$ anomaly are stretched. 
The muon $g-2$ anomaly can be explained within 1$\sigma$ (2$\sigma$) when $m_{\smu_1}\lesssim 390~(490)\GeV$ for $(m\w {L,R})_{\tilde \tau} = 1.5 \,(m\w {L,R})_{\tilde \ell}$ and when $m_{\smu_1}\lesssim 1.3~(1.6)\TeV$ for $(m\w {L,R})_{\tilde \tau} \gg (m\w {L,R})_{\tilde \ell}$.

Similarly to the universal slepton mass case, the neutralino relic abundance can be consistent with that of the dark matter, if the smuon mass is within $\sim10\%$ of the neutralino mass.
We checked that the masses can become as large as $\sim 1\TeV$ while keeping the sizable contribution to the muon $g-2$ as well as the correct relic abundance of the dark matter. 
Note that the dark matter scattering cross sections are tiny and it is challenging for the dark matter direct detection.  

Let us discuss future prospects in the above scenarios. 
For chargino-contribution dominated scenario, as mentioned in Ref.\cite{Endo:2020mqz}, future collider sensitivities may reach $m_{\charPM[1]} \lesssim 1.2\text{--}1.3\TeV$ by analyzing the NC/HW channel at the HL-LHC~\cite{ATL-PHYS-PUB-2018-048,CidVidal:2018eel}, 
and $m_{\charPM[1]} \lesssim 1.4\,(3.4)\TeV$ at a future $100\TeV$ $pp$ collider~\cite{Gori:2014oua,Bramante:2014tba}.
Besides, the electroweakinos will be able to be probed indirectly~\cite{Matsumoto:2017vfu,Matsumoto:2018ioi,Chigusa:2018vxz,Abe:2019egv}, \eg, for $m_{\charPM[1]} \lesssim 1.7\text{--}2.3\TeV$ at a $100\TeV$ $pp$ collider. 
Note that since these evaluations were performed on the simplified models, \ie, the sensitivities would be degraded in realistic setups.

Meanwhile, the ILC can play essential roles in testing the pure-bino-contribution dominated scenario.
In the universal slepton mass case, wide parameter regions that explain the muon $g-2$ discrepancy are accessible by the slepton searches if the collision energy is as large as $500\GeV$. The benchmark points BLR1 and BLR2 in Table~\ref{tab:BMP} can be probed even at ILC 250 GeV.
Furthermore, if the SUSY particles responsible for the muon $g-2$ are within the kinematical reach, it is possible to experimentally reconstruct the SUSY contribution to the muon $g-2$ by utilizing the precise ILC measurements~\cite{Endo:2013xka}.
It is noticed that the muon $g-2$ parameter space falls in the region in which sleptons and the LSP are degenerate.
Such spectra could also be a good target of the future LHC Runs, \eg, by investigating photon collisions~\cite{Beresford:2018pbt}.
Furthermore, higgsinos tend to be light when the lightest neutralino is light and/or $\tan\beta$ is large. 
Then, their production can be probed by analyzing events with SM bosons or taus with large missing transverse momentum. 
Such signals will be studied elsewhere.

In the scenario where the thermal relic bino dark matter abundance is consistent with the observed one, such as the benchmark points in Table~\ref{tab:BMP}, most of the viable parameter space will also be probed by the XENONnT experiment.
In addition, the muon $g-2$ parameter regions can be tested by measuring the branching ratio of the SM-like Higgs boson decay at future HL-LHC, ILC and/or FCC-ee/eh/hh especially when the lightest neutralino as well as the staus are light.

The Fermilab Muon $g-2$ collaboration has already completed Run-3 and the new results are anticipated with the full data set, which might 
shed further light on the SUSY scenarios.

%%%%%%%%%%%%%%%%%%%%%%%%%%%%%%%%%%%%%%%%%%%%%%%%%%
\section*{Acknowledgments}
%%%%%%%%%%%%%%%%%%%%%%%%%%%%%%%%%%%%%%%%%%%%%%%%%%
This work is supported in part by the Grant-in-Aid for Scientific Research on Innovative Areas 
(No.\,19H05810 [KH], No.\,19H05802 [KH]),
Scientific Research B (No.\,16H03991~[ME], No.\,20H01897 [KH]), and 
Early-Career Scientists (No.\,16K17681 [ME] and No.\,19K14706~[TK]).
The work of T.K.\ is also supported by the JSPS Core-to-Core Program, 
No.\,JPJSCCA20200002.
% SI has nothing to acknowledge

\bibliography{ref}

\providecommand{\href}[2]{#2}\begingroup\raggedright\begin{thebibliography}{100}

\bibitem{Aoyama:2020ynm}
T.~Aoyama {\em et~al.}, ``{The anomalous magnetic moment of the muon in the
  Standard Model},''
  \href{https://dx.doi.org/10.1016/j.physrep.2020.07.006}{Phys.\  Rept.\
  {\bfseries 887} (2020) 1--166} {\ttfamily
  [\href{https://arxiv.org/abs/2006.04822}{arXiv:2006.04822}]}.

\bibitem{Borsanyi:2020mff}
S.~Borsanyi {\em et~al.}, ``{Leading hadronic contribution to the muon 2
  magnetic moment from lattice QCD},''
  \href{https://dx.doi.org/10.1038/s41586-021-03418-1}{Nature (2021) }
  {\ttfamily [\href{https://arxiv.org/abs/2002.12347}{arXiv:2002.12347}]}.

\bibitem{Colangelo:2018mtw}
G.~Colangelo, M.~Hoferichter, and P.~Stoffer, ``{Two-pion contribution to
  hadronic vacuum polarization},''
  \href{https://dx.doi.org/10.1007/JHEP02(2019)006}{JHEP {\bfseries 02} (2019)
  006} {\ttfamily [\href{https://arxiv.org/abs/1810.00007}{arXiv:1810.00007}]}.

\bibitem{Hoferichter:2019mqg}
M.~Hoferichter, B.-L.~Hoid, and B.~Kubis, ``{Three-pion contribution to
  hadronic vacuum polarization},''
  \href{https://dx.doi.org/10.1007/JHEP08(2019)137}{JHEP {\bfseries 08} (2019)
  137} {\ttfamily [\href{https://arxiv.org/abs/1907.01556}{arXiv:1907.01556}]}.

\bibitem{Davier:2019can}
M.~Davier, A.~Hoecker, B.~Malaescu, and Z.~Zhang, ``{A new evaluation of the
  hadronic vacuum polarisation contributions to the muon anomalous magnetic
  moment and to $\alpha(m_Z^2)$},''
  \href{https://dx.doi.org/10.1140/epjc/s10052-020-7792-2}{Eur.\  Phys.\  J.\
  C {\bfseries 80} (2020) 241} {\ttfamily
  [\href{https://arxiv.org/abs/1908.00921}{arXiv:1908.00921}]} \textsl{[Erratum
  \href{https://doi.org/10.1140/epjc/s10052-020-7792-2}{ibid.\ {\bfseries 80}
  (2020) 410}]}.

\bibitem{Keshavarzi:2019abf}
A.~Keshavarzi, D.~Nomura, and T.~Teubner, ``{$g-2$ of charged leptons, $\alpha
  (M^2_Z)$ , and the hyperfine splitting of muonium},''
  \href{https://dx.doi.org/10.1103/PhysRevD.101.014029}{Phys.\  Rev.\  D
  {\bfseries 101} (2020) 014029} {\ttfamily
  [\href{https://arxiv.org/abs/1911.00367}{arXiv:1911.00367}]}.

\bibitem{Crivellin:2020zul}
A.~Crivellin, M.~Hoferichter, C.~A.~Manzari, and M.~Montull, ``{Hadronic Vacuum
  Polarization: $(g-2)_\mu$ versus Global Electroweak Fits},''
  \href{https://dx.doi.org/10.1103/PhysRevLett.125.091801}{Phys.\  Rev.\
  Lett.\  {\bfseries 125} (2020) 091801} {\ttfamily
  [\href{https://arxiv.org/abs/2003.04886}{arXiv:2003.04886}]}.

\bibitem{Keshavarzi:2020bfy}
A.~Keshavarzi, W.~J.~Marciano, M.~Passera, and A.~Sirlin, ``{Muon $g-2$ and
  $\Delta \alpha$ connection},''
  \href{https://dx.doi.org/10.1103/PhysRevD.102.033002}{Phys.\  Rev.\  D
  {\bfseries 102} (2020) 033002} {\ttfamily
  [\href{https://arxiv.org/abs/2006.12666}{arXiv:2006.12666}]}.

\bibitem{Malaescu:2020zuc}
B.~Malaescu and M.~Schott, ``{Impact of correlations between $a_{\mu }$ and
  $\alpha _\text {QED}$ on the EW fit},''
  \href{https://dx.doi.org/10.1140/epjc/s10052-021-08848-9}{Eur.\  Phys.\  J.\
  C {\bfseries 81} (2021) 46} {\ttfamily
  [\href{https://arxiv.org/abs/2008.08107}{arXiv:2008.08107}]}.

\bibitem{Colangelo:2020lcg}
G.~Colangelo, M.~Hoferichter, and P.~Stoffer, ``{Constraints on the two-pion
  contribution to hadronic vacuum polarization},''
  \href{https://dx.doi.org/10.1016/j.physletb.2021.136073}{Phys.\  Lett.\  B
  {\bfseries 814} (2021) 136073} {\ttfamily
  [\href{https://arxiv.org/abs/2010.07943}{arXiv:2010.07943}]}.

\bibitem{Bennett:2002jb}
{\bfseries Muon g-2} Collaboration, ``{Measurement of the positive muon
  anomalous magnetic moment to 0.7 ppm},''
  \href{https://dx.doi.org/10.1103/PhysRevLett.89.101804}{Phys.\  Rev.\  Lett.\
   {\bfseries 89} (2002) 101804}
{\ttfamily [\href{https://arxiv.org/abs/hep-ex/0208001}{hep-ex/0208001}]}.
%%CITATION = HEP-EX/0208001;%%.

\bibitem{Bennett:2004pv}
{\bfseries Muon g-2} Collaboration, ``{Measurement of the negative muon
  anomalous magnetic moment to 0.7 ppm},''
  \href{https://dx.doi.org/10.1103/PhysRevLett.92.161802}{Phys.\  Rev.\  Lett.\
   {\bfseries 92} (2004) 161802}
{\ttfamily [\href{https://arxiv.org/abs/hep-ex/0401008}{hep-ex/0401008}]}.
%%CITATION = HEP-EX/0401008;%%.

\bibitem{Bennett:2006fi}
{\bfseries Muon g-2} Collaboration, ``{Final Report of the Muon E821 Anomalous
  Magnetic Moment Measurement at BNL},''
  \href{https://dx.doi.org/10.1103/PhysRevD.73.072003}{Phys.\  Rev.\
  {\bfseries D73} (2006) 072003}
{\ttfamily [\href{https://arxiv.org/abs/hep-ex/0602035}{hep-ex/0602035}]}.
%%CITATION = HEP-EX/0602035;%%.

\bibitem{CODATA2018}
E.~Tiesinga, P.~J.~Mohr, D.~B.~Newell, and B.~N.~Taylor, ``{The 2018 CODATA
  Recommended Values of the Fundamental Physical Constants}.''
  \url{http://physics.nist.gov/constants} ({Web Version 8.0}), 2019.

\bibitem{Grange:2015fou}
{\bfseries Muon g-2} Collaboration, ``{Muon $g-2$ Technical Design Report}.''
{\ttfamily \href{https://arxiv.org/abs/1501.06858}{arXiv:1501.06858}}.
%%CITATION = ARXIV:1501.06858;%%.

\bibitem{Mibe:2011zz}
{\bfseries J-PARC g-2} Collaboration, ``{Measurement of muon $g-2$ and EDM with
  an ultra-cold muon beam at J-PARC},''
\href{https://dx.doi.org/10.1016/j.nuclphysbps.2011.06.039}{Nucl.\  Phys.\
  Proc.\  Suppl.\  {\bfseries 218} (2011) 242--246}.
%%CITATION = NUPHZ,218,242;%%.

\bibitem{Abe:2019thb}
M.~Abe {\em et~al.}, ``{A New Approach for Measuring the Muon Anomalous
  Magnetic Moment and Electric Dipole Moment},''
  \href{https://dx.doi.org/10.1093/ptep/ptz030}{PTEP {\bfseries 2019} (2019)
  053C02}
{\ttfamily [\href{https://arxiv.org/abs/1901.03047}{arXiv:1901.03047}]}.
%%CITATION = ARXIV:1901.03047;%%.

\bibitem{g-2Seminar20210407}
C.~Polly, on behalf of {\bfseries Muon g-2} Collaboration, ``First results from
  the Muon g-2 experiment at Fermilab,''
\newblock 7 Apr.\ 2021.
\newblock Seminar talk given at Fermilab.

\bibitem{Abi:2021gix}
{\bfseries Muon g-2} Collaboration, ``{Measurement of the Positive Muon
  Anomalous Magnetic Moment to 0.46 ppm},''
  \href{https://dx.doi.org/10.1103/PhysRevLett.126.141801}{Phys.\  Rev.\
  Lett.\  {\bfseries 126} (2021) 141801} {\ttfamily
  [\href{https://arxiv.org/abs/2104.03281}{arXiv:2104.03281}]}.

\bibitem{Chao:2021tvp}
E.-H.~Chao, {\em et al.}, ``{Hadronic light-by-light contribution to
  $(g-2)_\mu$ from lattice QCD: a complete calculation}.'' {\ttfamily
  \href{https://arxiv.org/abs/2104.02632}{arXiv:2104.02632}}.

\bibitem{Lopez:1993vi}
J.~L.~Lopez, D.~V.~Nanopoulos, and X.~Wang, ``{Large $(g-2)_{\mu}$ in SU(5)
  $\times$ U(1) Supergravity Models},''
  \href{https://dx.doi.org/10.1103/PhysRevD.49.366}{Phys.\  Rev.\  {\bfseries
  D49} (1994) 366--372}
{\ttfamily [\href{https://arxiv.org/abs/hep-ph/9308336}{hep-ph/9308336}]}.
%%CITATION = HEP-PH/9308336;%%.

\bibitem{Chattopadhyay:1995ae}
U.~Chattopadhyay and P.~Nath, ``{Probing Supergravity Grand Unification in the
  Brookhaven $g-2$ Experiment},''
  \href{https://dx.doi.org/10.1103/PhysRevD.53.1648}{Phys.\  Rev.\  {\bfseries
  D53} (1996) 1648--1657}
{\ttfamily [\href{https://arxiv.org/abs/hep-ph/9507386}{hep-ph/9507386}]}.
%%CITATION = HEP-PH/9507386;%%.

\bibitem{Moroi:1995yh}
T.~Moroi, ``{The Muon anomalous magnetic dipole moment in the minimal
  supersymmetric standard model},''
  \href{https://dx.doi.org/10.1103/PhysRevD.53.6565,
  10.1103/PhysRevD.56.4424}{Phys.\  Rev.\  {\bfseries D53} (1996) 6565--6575}
  {\ttfamily [\href{https://arxiv.org/abs/hep-ph/9512396}{hep-ph/9512396}]}
\textsl{[Erratum \href{https://doi.org/10.1103/PhysRevD.56.4424}{ibid.\
  {\bfseries D56} (1997) 4424}]}.
%%CITATION = HEP-PH/9512396;%%.

\bibitem{Zhu:2016ncq}
B.~Zhu, R.~Ding, and T.~Li, ``{Higgs mass and muon anomalous magnetic moment in
  the MSSM with gauge-gravity hybrid mediation},''
  \href{https://dx.doi.org/10.1103/PhysRevD.96.035029}{Phys.\  Rev.\
  {\bfseries D96} (2017) 035029}
{\ttfamily [\href{https://arxiv.org/abs/1610.09840}{arXiv:1610.09840}]}.
%%CITATION = ARXIV:1610.09840;%%.

\bibitem{Choudhury:2017fuu}
A.~Choudhury, L.~Darmé, L.~Roszkowski, E.~M.~Sessolo, and S.~Trojanowski,
  ``{Muon $g-2$ and related phenomenology in constrained vector-like extensions
  of the MSSM},'' \href{https://dx.doi.org/10.1007/JHEP05(2017)072}{JHEP
  {\bfseries 05} (2017) 072}
{\ttfamily [\href{https://arxiv.org/abs/1701.08778}{arXiv:1701.08778}]}.
%%CITATION = ARXIV:1701.08778;%%.

\bibitem{Yanagida:2017dao}
T.~T.~Yanagida and N.~Yokozaki, ``{Muon $g - 2$ in MSSM gauge mediation
  revisited},'' \href{https://dx.doi.org/10.1016/j.physletb.2017.07.002}{Phys.\
   Lett.\  {\bfseries B772} (2017) 409--414}
{\ttfamily [\href{https://arxiv.org/abs/1704.00711}{arXiv:1704.00711}]}.
%%CITATION = ARXIV:1704.00711;%%.

\bibitem{Endo:2017zrj}
M.~Endo, K.~Hamaguchi, S.~Iwamoto, and K.~Yanagi, ``{Probing minimal SUSY
  scenarios in the light of muon $g-2$ and dark matter},''
  \href{https://dx.doi.org/10.1007/JHEP06(2017)031}{JHEP {\bfseries 06} (2017)
  031}
{\ttfamily [\href{https://arxiv.org/abs/1704.05287}{arXiv:1704.05287}]}.
%%CITATION = ARXIV:1704.05287;%%.

\bibitem{Hagiwara:2017lse}
K.~Hagiwara, K.~Ma, and S.~Mukhopadhyay, ``{Closing in on the chargino
  contribution to the muon $g-2$ in the MSSM: current LHC constraints},''
  \href{https://dx.doi.org/10.1103/PhysRevD.97.055035}{Phys.\  Rev.\
  {\bfseries D97} (2018) 055035}
{\ttfamily [\href{https://arxiv.org/abs/1706.09313}{arXiv:1706.09313}]}.
%%CITATION = ARXIV:1706.09313;%%.

\bibitem{Chakraborti:2017vxz}
M.~Chakraborti, A.~Datta, N.~Ganguly, and S.~Poddar, ``{Multilepton signals of
  heavier electroweakinos at the LHC},''
  \href{https://dx.doi.org/10.1007/JHEP11(2017)117}{JHEP {\bfseries 11} (2017)
  117}
{\ttfamily [\href{https://arxiv.org/abs/1707.04410}{arXiv:1707.04410}]}.
%%CITATION = ARXIV:1707.04410;%%.

\bibitem{Choudhury:2017acn}
A.~Choudhury, S.~Rao, and L.~Roszkowski, ``{Impact of LHC data on muon $g-2$
  solutions in a vectorlike extension of the constrained MSSM},''
  \href{https://dx.doi.org/10.1103/PhysRevD.96.075046}{Phys.\  Rev.\
  {\bfseries D96} (2017) 075046}
{\ttfamily [\href{https://arxiv.org/abs/1708.05675}{arXiv:1708.05675}]}.
%%CITATION = ARXIV:1708.05675;%%.

\bibitem{Ajaib:2017zba}
M.~A.~Ajaib, ``{SU(5) with nonuniversal gaugino masses},''
  \href{https://dx.doi.org/10.1142/S0217751X1850032X}{Int.\  J.\  Mod.\  Phys.\
   {\bfseries A33} (2018) 1850032}
{\ttfamily [\href{https://arxiv.org/abs/1711.02560}{arXiv:1711.02560}]}.
%%CITATION = ARXIV:1711.02560;%%.

\bibitem{Belyaev:2018vkl}
A.~S.~Belyaev, S.~F.~King, and P.~B.~Schaefers, ``{Muon g-2 and dark matter
  suggest nonuniversal gaugino masses: $\mathrm{SU}(5)\times A_4$ case study at
  the LHC},'' \href{https://dx.doi.org/10.1103/PhysRevD.97.115002}{Phys.\
  Rev.\  {\bfseries D97} (2018) 115002}
{\ttfamily [\href{https://arxiv.org/abs/1801.00514}{arXiv:1801.00514}]}.
%%CITATION = ARXIV:1801.00514;%%.

\bibitem{Bhattacharyya:2018inr}
G.~Bhattacharyya, T.~T.~Yanagida, and N.~Yokozaki, ``{An extended gauge
  mediation for muon $(g-2)$ explanation},''
  \href{https://dx.doi.org/10.1016/j.physletb.2018.07.037}{Phys.\  Lett.\
  {\bfseries B784} (2018) 118--121}
{\ttfamily [\href{https://arxiv.org/abs/1805.01607}{arXiv:1805.01607}]}.
%%CITATION = ARXIV:1805.01607;%%.

\bibitem{Abel:2018ekz}
S.~Abel, D.~G.~Cerdeño, and S.~Robles, ``{The Power of Genetic Algorithms:
  what remains of the pMSSM?}''
{\ttfamily \href{https://arxiv.org/abs/1805.03615}{arXiv:1805.03615}}.
%%CITATION = ARXIV:1805.03615;%%.

\bibitem{Cao:2018rix}
J.~Cao, Y.~He, L.~Shang, Y.~Zhang, and P.~Zhu, ``{Current status of a natural
  NMSSM in light of LHC 13 TeV data and XENON-1T results},''
  \href{https://dx.doi.org/10.1103/PhysRevD.99.075020}{Phys.\  Rev.\
  {\bfseries D99} (2019) 075020}
{\ttfamily [\href{https://arxiv.org/abs/1810.09143}{arXiv:1810.09143}]}.
%%CITATION = ARXIV:1810.09143;%%.

\bibitem{Dutta:2018fge}
B.~Dutta and Y.~Mimura, ``{Electron $g-2$ with flavor violation in MSSM},''
  \href{https://dx.doi.org/10.1016/j.physletb.2018.12.070}{Phys.\  Lett.\
  {\bfseries B790} (2019) 563--567}
{\ttfamily [\href{https://arxiv.org/abs/1811.10209}{arXiv:1811.10209}]}.
%%CITATION = ARXIV:1811.10209;%%.

\bibitem{Cox:2018vsv}
P.~Cox, C.~Han, T.~T.~Yanagida, and N.~Yokozaki, ``{Gaugino mediation scenarios
  for muon $g-2$ and dark matter},''
  \href{https://dx.doi.org/10.1007/JHEP08(2019)097}{JHEP {\bfseries 08} (2019)
  097}
{\ttfamily [\href{https://arxiv.org/abs/1811.12699}{arXiv:1811.12699}]}.
%%CITATION = ARXIV:1811.12699;%%.

\bibitem{Tran:2018kxv}
H.~M.~Tran and H.~T.~Nguyen, ``{GUT-inspired MSSM in light of muon $g-2$ and
  LHC results at $\sqrt{s}=13$ TeV},''
  \href{https://dx.doi.org/10.1103/PhysRevD.99.035040}{Phys.\  Rev.\
  {\bfseries D99} (2019) 035040}
{\ttfamily [\href{https://arxiv.org/abs/1812.11757}{arXiv:1812.11757}]}.
%%CITATION = ARXIV:1812.11757;%%.

\bibitem{Ibe:2019jbx}
M.~Ibe, M.~Suzuki, T.~T.~Yanagida, and N.~Yokozaki, ``{Muon $g-2$ in
  Split-Family SUSY in light of LHC Run II},''
  \href{https://dx.doi.org/10.1140/epjc/s10052-019-7186-5}{Eur.\  Phys.\  J.\
  {\bfseries C79} (2019) 688}
{\ttfamily [\href{https://arxiv.org/abs/1903.12433}{arXiv:1903.12433}]}.
%%CITATION = ARXIV:1903.12433;%%.

\bibitem{Badziak:2019gaf}
M.~Badziak and K.~Sakurai, ``{Explanation of electron and muon $g - 2$
  anomalies in the MSSM},''
  \href{https://dx.doi.org/10.1007/JHEP10(2019)024}{JHEP {\bfseries 10} (2019)
  024}
{\ttfamily [\href{https://arxiv.org/abs/1908.03607}{arXiv:1908.03607}]}.
%%CITATION = ARXIV:1908.03607;%%.

\bibitem{Abdughani:2019wai}
M.~Abdughani, K.-I.~Hikasa, L.~Wu, J.~M.~Yang, and J.~Zhao, ``{Testing
  electroweak SUSY for muon $g-2$ and dark matter at the LHC and beyond},''
  \href{https://dx.doi.org/10.1007/JHEP11(2019)095}{JHEP {\bfseries 11} (2019)
  095}
{\ttfamily [\href{https://arxiv.org/abs/1909.07792}{arXiv:1909.07792}]}.
%%CITATION = ARXIV:1909.07792;%%.

\bibitem{Kpatcha:2019pve}
E.~Kpatcha, I.~Lara, D.~E.~López-Fogliani, C.~Muñoz, and N.~Nagata,
  ``{Explaining muon $g-2$ data in the $\mu\nu$SSM},''
  \href{https://dx.doi.org/10.1140/epjc/s10052-021-08938-8}{Eur.\  Phys.\  J.\
  {\bfseries C81} (2021) 154}
{\ttfamily [\href{https://arxiv.org/abs/1912.04163}{arXiv:1912.04163}]}.
%%CITATION = ARXIV:1912.04163;%%.

\bibitem{Yanagida:2020jzy}
T.~T.~Yanagida, W.~Yin, and N.~Yokozaki, ``{Muon $g-2$ in Higgs-anomaly
  mediation},'' \href{https://dx.doi.org/10.1007/JHEP06(2020)154}{JHEP
  {\bfseries 06} (2020) 154} {\ttfamily
  [\href{https://arxiv.org/abs/2001.02672}{arXiv:2001.02672}]}.

\bibitem{Han:2020exx}
C.~Han, {\em et al.}, ``{LFV and (g-2) in non-universal SUSY models with light
  higgsinos},'' \href{https://dx.doi.org/10.1007/JHEP05(2020)102}{JHEP
  {\bfseries 05} (2020) 102}
{\ttfamily [\href{https://arxiv.org/abs/2003.06187}{arXiv:2003.06187}]}.
%%CITATION = ARXIV:2003.06187;%%.

\bibitem{Chakraborti:2020vjp}
M.~Chakraborti, S.~Heinemeyer, and I.~Saha, ``{Improved $(g-2)_\mu$
  Measurements and Supersymmetry},''
  \href{https://dx.doi.org/10.1140/epjc/s10052-020-08504-8}{Eur.\  Phys.\  J.\
  {\bfseries C80} (2020) 984}
{\ttfamily [\href{https://arxiv.org/abs/2006.15157}{arXiv:2006.15157}]}.
%%CITATION = ARXIV:2006.15157;%%.

\bibitem{Nagai:2020xbq}
R.~Nagai and N.~Yokozaki, ``{Lepton flavor violations in SUSY models for muon
  $g-2$ with right-handed neutrinos},''
  \href{https://dx.doi.org/10.1007/JHEP01(2021)099}{JHEP {\bfseries 01} (2021)
  099}
{\ttfamily [\href{https://arxiv.org/abs/2007.00943}{arXiv:2007.00943}]}.
%%CITATION = ARXIV:2007.00943;%%.

\bibitem{Chakraborti:2021kkr}
M.~Chakraborti, S.~Heinemeyer, and I.~Saha, ``{Improved $(g-2)_\mu$
  Measurements and Wino/Higgsino Dark Matter}.''
{\ttfamily \href{https://arxiv.org/abs/2103.13403}{arXiv:2103.13403}}.
%%CITATION = ARXIV:2103.13403;%%.

\bibitem{Endo:2020mqz}
M.~Endo, K.~Hamaguchi, S.~Iwamoto, and T.~Kitahara, ``{Muon $g-2$ vs LHC Run 2
  in supersymmetric models},''
  \href{https://dx.doi.org/10.1007/JHEP04(2020)165}{JHEP {\bfseries 04} (2020)
  165} {\ttfamily [\href{https://arxiv.org/abs/2001.11025}{arXiv:2001.11025}]}.

\bibitem{Aad:2020qnn}
{\bfseries ATLAS} Collaboration, ``{Search for direct production of
  electroweakinos in final states with missing transverse momentum and a Higgs
  boson decaying into photons in pp collisions at $ \sqrt{s} $ = 13 TeV with
  the ATLAS detector},'' \href{https://dx.doi.org/10.1007/JHEP10(2020)005}{JHEP
  {\bfseries 10} (2020) 005} {\ttfamily
  [\href{https://arxiv.org/abs/2004.10894}{arXiv:2004.10894}]}.

\bibitem{Sirunyan:2020eab}
{\bfseries CMS} Collaboration, ``{Search for supersymmetry in final states with
  two oppositely charged same-flavor leptons and missing transverse momentum in
  proton-proton collisions at $\sqrt{s}=$ 13 TeV}.'' {\ttfamily
  \href{https://arxiv.org/abs/2012.08600}{arXiv:2012.08600}}.

\bibitem{Endo:2013lva}
M.~Endo, K.~Hamaguchi, T.~Kitahara, and T.~Yoshinaga, ``{Probing Bino
  contribution to muon $g - 2$},''
  \href{https://dx.doi.org/10.1007/JHEP11(2013)013}{JHEP {\bfseries 11} (2013)
  013}
{\ttfamily [\href{https://arxiv.org/abs/1309.3065}{arXiv:1309.3065}]}.
%%CITATION = ARXIV:1309.3065;%%.

\bibitem{Iwamoto:2021aaf}
S.~Iwamoto, T.~T.~Yanagida, and N.~Yokozaki, ``{Wino-Higgsino dark matter in
  the MSSM from the $g-2$ anomaly}.'' {\ttfamily
  \href{https://arxiv.org/abs/2104.03223}{arXiv:2104.03223}}.

\bibitem{Muhlleitner:2003vg}
M.~Muhlleitner, A.~Djouadi, and Y.~Mambrini, ``{SDECAY: A Fortran code for the
  decays of the supersymmetric particles in the MSSM},''
  \href{https://dx.doi.org/10.1016/j.cpc.2005.01.012}{Comput.\  Phys.\
  Commun.\  {\bfseries 168} (2005) 46--70}
{\ttfamily [\href{https://arxiv.org/abs/hep-ph/0311167}{hep-ph/0311167}]}.
%%CITATION = HEP-PH/0311167;%%.

\bibitem{Djouadi:2006bz}
M.~M.~M{\"u}hlleitner, A.~Djouadi, and M.~Spira, ``{Decays of Supersymmetric
  Particles --- the Program SUSY-HIT},'' Acta Phys.\  Polon.\  {\bfseries B38}
  (2007) 635--644
{\ttfamily [\href{https://arxiv.org/abs/hep-ph/0609292}{hep-ph/0609292}]}.
%%CITATION = HEP-PH/0609292;%%.

\bibitem{Athron:2015rva}
P.~Athron, {\em et al.}, ``{Gm2Calc: Precise MSSM Prediction for $(g - 2)$ of
  the Muon},'' \href{https://dx.doi.org/10.1140/epjc/s10052-015-3870-2}{Eur.\
  Phys.\  J.\  {\bfseries C76} (2016) 62}
{\ttfamily [\href{https://arxiv.org/abs/1510.08071}{arXiv:1510.08071}]}.
%%CITATION = ARXIV:1510.08071;%%.

\bibitem{Aad:2019vnb}
{\bfseries ATLAS} Collaboration, ``{Search for electroweak production of
  charginos and sleptons decaying into final states with two leptons and
  missing transverse momentum in $\sqrt{s}=13$ TeV $pp$ collisions using the
  ATLAS detector},''
  \href{https://dx.doi.org/10.1140/epjc/s10052-019-7594-6}{Eur.\  Phys.\  J.\
  C {\bfseries 80} (2020) 123} {\ttfamily
  [\href{https://arxiv.org/abs/1908.08215}{arXiv:1908.08215}]}.

\bibitem{Aad:2019vvf}
{\bfseries ATLAS} Collaboration, ``{Search for direct production of
  electroweakinos in final states with one lepton, missing transverse momentum
  and a Higgs boson decaying into two $b$-jets in $pp$ collisions at
  $\sqrt{s}=13$ TeV with the ATLAS detector},''
  \href{https://dx.doi.org/10.1140/epjc/s10052-020-8050-3}{Eur.\  Phys.\  J.\
  C {\bfseries 80} (2020) 691} {\ttfamily
  [\href{https://arxiv.org/abs/1909.09226}{arXiv:1909.09226}]}.

\bibitem{Sirunyan:2017lae}
{\bfseries CMS} Collaboration, ``{Search for electroweak production of
  charginos and neutralinos in multilepton final states in proton-proton
  collisions at $\sqrt{s}=$ 13 TeV},''
  \href{https://dx.doi.org/10.1007/JHEP03(2018)166}{JHEP {\bfseries 03} (2018)
  166}
{\ttfamily [\href{https://arxiv.org/abs/1709.05406}{arXiv:1709.05406}]}.
%%CITATION = ARXIV:1709.05406;%%.

\bibitem{Aaboud:2018jiw}
{\bfseries ATLAS} Collaboration, ``{Search for electroweak production of
  supersymmetric particles in final states with two or three leptons at
  $\sqrt{s}=13\,$TeV with the ATLAS detector},''
  \href{https://dx.doi.org/10.1140/epjc/s10052-018-6423-7}{Eur.\  Phys.\  J.\
  {\bfseries C78} (2018) 995}
{\ttfamily [\href{https://arxiv.org/abs/1803.02762}{arXiv:1803.02762}]}.
%%CITATION = ARXIV:1803.02762;%%.

\bibitem{Endo:2013bba}
M.~Endo, K.~Hamaguchi, S.~Iwamoto, and T.~Yoshinaga, ``{Muon $g-2$ vs LHC in
  supersymmetric models},''
  \href{https://dx.doi.org/10.1007/JHEP01(2014)123}{JHEP {\bfseries 01} (2014)
  123}
{\ttfamily [\href{https://arxiv.org/abs/1303.4256}{arXiv:1303.4256}]}.
%%CITATION = ARXIV:1303.4256;%%.

\bibitem{ATLAS-CONF-2020-015}
{\bfseries ATLAS} Collaboration, ``{Search for chargino-neutralino pair
  production in final states with three leptons and missing transverse momentum
  in $\sqrt{s}$=13TeV $pp$ collisions with the ATLAS detector},''
  \href{http://cds.cern.ch/record/2719521}{ATLAS--CONF--2020--015}, CERN, 2020.

\bibitem{CMS-PAS-SUS-19-012}
{\bfseries CMS} Collaboration, ``{Search for electroweak production of
  charginos and neutralinos in proton-proton collisions at
  $\sqrt{s}=13$\,TeV},''
  \href{https://cds.cern.ch/record/2752640}{CMS--PAS--SUS--19--012}, CERN,
  2021.

\bibitem{CMS-PAS-SUS-20-003}
{\bfseries CMS} Collaboration, ``{Search for chargino-neutralino production in
  final states with a Higgs boson and a W boson},''
  \href{https://cds.cern.ch/record/2758360/}{CMS--PAS--SUS--20--003}, CERN,
  2021.

\bibitem{vonWeitershausen:2010zr}
P.~von Weitershausen, M.~Schafer, H.~Stockinger-Kim, and D.~Stockinger,
  ``{Photonic SUSY Two-Loop Corrections to the Muon Magnetic Moment},''
  \href{https://dx.doi.org/10.1103/PhysRevD.81.093004}{Phys.\  Rev.\  D
  {\bfseries 81} (2010) 093004} {\ttfamily
  [\href{https://arxiv.org/abs/1003.5820}{arXiv:1003.5820}]}.

\bibitem{Marchetti:2008hw}
S.~Marchetti, S.~Mertens, U.~Nierste, and D.~Stockinger, ``{Tan(beta)-enhanced
  supersymmetric corrections to the anomalous magnetic moment of the muon},''
  \href{https://dx.doi.org/10.1103/PhysRevD.79.013010}{Phys.\  Rev.\  D
  {\bfseries 79} (2009) 013010} {\ttfamily
  [\href{https://arxiv.org/abs/0808.1530}{arXiv:0808.1530}]}.

\bibitem{Girrbach:2009uy}
J.~Girrbach, S.~Mertens, U.~Nierste, and S.~Wiesenfeldt, ``{Lepton flavour
  violation in the MSSM},''
  \href{https://dx.doi.org/10.1007/JHEP05(2010)026}{JHEP {\bfseries 05} (2010)
  026} {\ttfamily [\href{https://arxiv.org/abs/0910.2663}{arXiv:0910.2663}]}.

\bibitem{Nojiri:1996fp}
M.~M.~Nojiri, K.~Fujii, and T.~Tsukamoto, ``{Confronting the minimal
  supersymmetric standard model with the study of scalar leptons at future
  linear e+ e- colliders},''
  \href{https://dx.doi.org/10.1103/PhysRevD.54.6756}{Phys.\  Rev.\  D
  {\bfseries 54} (1996) 6756--6776} {\ttfamily
  [\href{https://arxiv.org/abs/hep-ph/9606370}{hep-ph/9606370}]}.

\bibitem{Nojiri:1997ma}
M.~M.~Nojiri, D.~M.~Pierce, and Y.~Yamada, ``{Slepton production as a probe of
  the squark mass scale},''
  \href{https://dx.doi.org/10.1103/PhysRevD.57.1539}{Phys.\  Rev.\  D
  {\bfseries 57} (1998) 1539--1552} {\ttfamily
  [\href{https://arxiv.org/abs/hep-ph/9707244}{hep-ph/9707244}]}.

\bibitem{Cheng:1997sq}
H.-C.~Cheng, J.~L.~Feng, and N.~Polonsky, ``{Superoblique corrections and
  nondecoupling of supersymmetry breaking},''
  \href{https://dx.doi.org/10.1103/PhysRevD.56.6875}{Phys.\  Rev.\  D
  {\bfseries 56} (1997) 6875--6884} {\ttfamily
  [\href{https://arxiv.org/abs/hep-ph/9706438}{hep-ph/9706438}]}.

\bibitem{Cheng:1997vy}
H.-C.~Cheng, J.~L.~Feng, and N.~Polonsky, ``{Signatures of multi - TeV scale
  particles in supersymmetric theories},''
  \href{https://dx.doi.org/10.1103/PhysRevD.57.152}{Phys.\  Rev.\  D {\bfseries
  57} (1998) 152--169} {\ttfamily
  [\href{https://arxiv.org/abs/hep-ph/9706476}{hep-ph/9706476}]}.

\bibitem{Katz:1998br}
E.~Katz, L.~Randall, and S.-f.~Su, ``{Supersymmetric partners of oblique
  corrections},''
  \href{https://dx.doi.org/10.1016/S0550-3213(98)00632-4}{Nucl.\  Phys.\  B
  {\bfseries 536} (1998) 3--28} {\ttfamily
  [\href{https://arxiv.org/abs/hep-ph/9801416}{hep-ph/9801416}]}.

\bibitem{Fargnoli:2013zda}
H.~G.~Fargnoli, C.~Gnendiger, S.~Paßehr, D.~Stöckinger, and
  H.~Stöckinger-Kim, ``{Non-decoupling two-loop corrections to $(g-2)_\mu$
  from fermion/sfermion loops in the MSSM},''
  \href{https://dx.doi.org/10.1016/j.physletb.2013.09.034}{Phys.\  Lett.\
  {\bfseries B726} (2013) 717--724}
{\ttfamily [\href{https://arxiv.org/abs/1309.0980}{arXiv:1309.0980}]}.
%%CITATION = ARXIV:1309.0980;%%.

\bibitem{Aad:2014vma}
{\bfseries ATLAS} Collaboration, ``{Search for direct production of charginos,
  neutralinos and sleptons in final states with two leptons and missing
  transverse momentum in $pp$ collisions at $\sqrt{s} =$ 8 TeV with the ATLAS
  detector},'' \href{https://dx.doi.org/10.1007/JHEP05(2014)071}{JHEP
  {\bfseries 05} (2014) 071} {\ttfamily
  [\href{https://arxiv.org/abs/1403.5294}{arXiv:1403.5294}]}.

\bibitem{Aad:2019qnd}
{\bfseries ATLAS} Collaboration, ``{Searches for electroweak production of
  supersymmetric particles with compressed mass spectra in $\sqrt{s}=$ 13 TeV
  $pp$ collisions with the ATLAS detector},''
  \href{https://dx.doi.org/10.1103/PhysRevD.101.052005}{Phys.\  Rev.\  D
  {\bfseries 101} (2020) 052005} {\ttfamily
  [\href{https://arxiv.org/abs/1911.12606}{arXiv:1911.12606}]}.

\bibitem{LEP2SUSYWG:04-01.1}
{LEP2 SUSY Working Group (ALEPH, DELPHI, L3, OPAL)}, ``{Combined LEP
  Selectron/Smuon/Stau Results, 183--208 GeV}.'' {Notes LEPSUSYWG/04--01.1}.
\newblock \url{http://lepsusy.web.cern.ch/lepsusy/Welcome.html}.

\bibitem{Aad:2019byo}
{\bfseries ATLAS} Collaboration, ``{Search for direct stau production in events
  with two hadronic $\tau$-leptons in $\sqrt{s} = 13$ TeV $pp$ collisions with
  the ATLAS detector},''
  \href{https://dx.doi.org/10.1103/PhysRevD.101.032009}{Phys.\  Rev.\  D
  {\bfseries 101} (2020) 032009} {\ttfamily
  [\href{https://arxiv.org/abs/1911.06660}{arXiv:1911.06660}]}.

\bibitem{CMS:2019eln}
{\bfseries CMS} Collaboration, ``{Search for direct pair production of
  supersymmetric partners to the $\tau$ lepton in proton-proton collisions at
  $\sqrt{s}=$ 13 TeV},''
  \href{https://dx.doi.org/10.1140/epjc/s10052-020-7739-7}{Eur.\  Phys.\  J.\
  C {\bfseries 80} (2020) 189} {\ttfamily
  [\href{https://arxiv.org/abs/1907.13179}{arXiv:1907.13179}]}.

\bibitem{LEP2SUSYWG:01-03.1}
{LEP2 SUSY Working Group (ALEPH, DELPHI, L3, OPAL)}, ``{Combined LEP Chargino
  Results, up to 208 GeV for large m0}.'' {Notes LEPSUSYWG/01--03.1}.
\newblock \url{http://lepsusy.web.cern.ch/lepsusy/Welcome.html}.

\bibitem{Carena:2011aa}
M.~Carena, S.~Gori, N.~R.~Shah, and C.~E.~M.~Wagner, ``{A 125 GeV SM-like Higgs
  in the MSSM and the $\gamma \gamma$ rate},''
  \href{https://dx.doi.org/10.1007/JHEP03(2012)014}{JHEP {\bfseries 03} (2012)
  014} {\ttfamily [\href{https://arxiv.org/abs/1112.3336}{arXiv:1112.3336}]}.

\bibitem{Carena:2012gp}
M.~Carena, S.~Gori, N.~R.~Shah, C.~E.~M.~Wagner, and L.-T.~Wang, ``{Light Stau
  Phenomenology and the Higgs $\gamma\gamma$ Rate},''
  \href{https://dx.doi.org/10.1007/JHEP07(2012)175}{JHEP {\bfseries 07} (2012)
  175} {\ttfamily [\href{https://arxiv.org/abs/1205.5842}{arXiv:1205.5842}]}.

\bibitem{Kitahara:2012pb}
T.~Kitahara, ``{Vacuum Stability Constraints on the Enhancement of the $h \to
  \gamma \gamma$ rate in the MSSM},''
  \href{https://dx.doi.org/10.1007/JHEP11(2012)021}{JHEP {\bfseries 11} (2012)
  021} {\ttfamily [\href{https://arxiv.org/abs/1208.4792}{arXiv:1208.4792}]}.

\bibitem{Carena:2012mw}
M.~Carena, S.~Gori, I.~Low, N.~R.~Shah, and C.~E.~M.~Wagner, ``{Vacuum
  Stability and Higgs Diphoton Decays in the MSSM},''
  \href{https://dx.doi.org/10.1007/JHEP02(2013)114}{JHEP {\bfseries 02} (2013)
  114} {\ttfamily [\href{https://arxiv.org/abs/1211.6136}{arXiv:1211.6136}]}.

\bibitem{Zyla:2020zbs}
{\bfseries Particle Data Group} Collaboration, ``{Review of Particle
  Physics},'' \href{https://dx.doi.org/10.1093/ptep/ptaa104}{PTEP {\bfseries
  2020} (2020) 083C01}.

\bibitem{deBlas:2019rxi}
J.~de~Blas {\em et~al.}, ``{Higgs Boson Studies at Future Particle
  Colliders},'' \href{https://dx.doi.org/10.1007/JHEP01(2020)139}{JHEP
  {\bfseries 01} (2020) 139} {\ttfamily
  [\href{https://arxiv.org/abs/1905.03764}{arXiv:1905.03764}]}.

\bibitem{Aghanim:2018eyx}
{\bfseries Planck} Collaboration, ``{Planck 2018 results. VI. Cosmological
  parameters},'' \href{https://dx.doi.org/10.1051/0004-6361/201833910}{Astron.\
   Astrophys.\  {\bfseries 641} (2020) A6} {\ttfamily
  [\href{https://arxiv.org/abs/1807.06209}{arXiv:1807.06209}]}.

\bibitem{Belanger:2013oya}
G.~Belanger, F.~Boudjema, A.~Pukhov, and A.~Semenov, ``{micrOMEGAs$\_$3: A
  program for calculating dark matter observables},''
  \href{https://dx.doi.org/10.1016/j.cpc.2013.10.016}{Comput.\  Phys.\
  Commun.\  {\bfseries 185} (2014) 960--985} {\ttfamily
  [\href{https://arxiv.org/abs/1305.0237}{arXiv:1305.0237}]}.

\bibitem{Belanger:2010pz}
G.~Belanger, F.~Boudjema, A.~Pukhov, and A.~Semenov, ``{micrOMEGAs: A Tool for
  dark matter studies},''
  \href{https://dx.doi.org/10.1393/ncc/i2010-10591-3}{Nuovo Cim.\  C {\bfseries
  033N2} (2010) 111--116} {\ttfamily
  [\href{https://arxiv.org/abs/1005.4133}{arXiv:1005.4133}]}.

\bibitem{Belanger:2008sj}
G.~Belanger, F.~Boudjema, A.~Pukhov, and A.~Semenov, ``{Dark matter direct
  detection rate in a generic model with micrOMEGAs 2.2},''
  \href{https://dx.doi.org/10.1016/j.cpc.2008.11.019}{Comput.\  Phys.\
  Commun.\  {\bfseries 180} (2009) 747--767} {\ttfamily
  [\href{https://arxiv.org/abs/0803.2360}{arXiv:0803.2360}]}.

\bibitem{Belanger:2006is}
G.~Belanger, F.~Boudjema, A.~Pukhov, and A.~Semenov, ``{MicrOMEGAs 2.0: A
  Program to calculate the relic density of dark matter in a generic model},''
  \href{https://dx.doi.org/10.1016/j.cpc.2006.11.008}{Comput.\  Phys.\
  Commun.\  {\bfseries 176} (2007) 367--382} {\ttfamily
  [\href{https://arxiv.org/abs/hep-ph/0607059}{hep-ph/0607059}]}.

\bibitem{Akerib:2016vxi}
{\bfseries LUX} Collaboration, ``{Results from a search for dark matter in the
  complete LUX exposure},''
  \href{https://dx.doi.org/10.1103/PhysRevLett.118.021303}{Phys.\  Rev.\
  Lett.\  {\bfseries 118} (2017) 021303} {\ttfamily
  [\href{https://arxiv.org/abs/1608.07648}{arXiv:1608.07648}]}.

\bibitem{Cui:2017nnn}
{\bfseries PandaX-II} Collaboration, ``{Dark Matter Results From 54-Ton-Day
  Exposure of PandaX-II Experiment},''
  \href{https://dx.doi.org/10.1103/PhysRevLett.119.181302}{Phys.\  Rev.\
  Lett.\  {\bfseries 119} (2017) 181302} {\ttfamily
  [\href{https://arxiv.org/abs/1708.06917}{arXiv:1708.06917}]}.

\bibitem{Aprile:2018dbl}
{\bfseries XENON} Collaboration, ``{Dark Matter Search Results from a One
  Ton-Year Exposure of XENON1T},''
  \href{https://dx.doi.org/10.1103/PhysRevLett.121.111302}{Phys.\  Rev.\
  Lett.\  {\bfseries 121} (2018) 111302} {\ttfamily
  [\href{https://arxiv.org/abs/1805.12562}{arXiv:1805.12562}]}.

\bibitem{Kitahara:2013lfa}
T.~Kitahara and T.~Yoshinaga, ``{Stau with Large Mass Difference and
  Enhancement of the Higgs to Diphoton Decay Rate in the MSSM},''
  \href{https://dx.doi.org/10.1007/JHEP05(2013)035}{JHEP {\bfseries 05} (2013)
  035} {\ttfamily [\href{https://arxiv.org/abs/1303.0461}{arXiv:1303.0461}]}.

\bibitem{Coleman:1977py}
S.~R.~Coleman, ``{The Fate of the False Vacuum. 1. Semiclassical Theory},''
  \href{https://dx.doi.org/10.1103/PhysRevD.16.1248}{Phys.\  Rev.\  D
  {\bfseries 15} (1977) 2929--2936}. [Erratum: Phys.Rev.D 16, 1248 (1977)].

\bibitem{Wainwright:2011kj}
C.~L.~Wainwright, ``{CosmoTransitions: Computing Cosmological Phase Transition
  Temperatures and Bubble Profiles with Multiple Fields},''
  \href{https://dx.doi.org/10.1016/j.cpc.2012.04.004}{Comput.\  Phys.\
  Commun.\  {\bfseries 183} (2012) 2006--2013} {\ttfamily
  [\href{https://arxiv.org/abs/1109.4189}{arXiv:1109.4189}]}.

\bibitem{Endo:2010ya}
M.~Endo, K.~Hamaguchi, and K.~Nakaji, ``{Probing High Reheating Temperature
  Scenarios at the LHC with Long-Lived Staus},''
  \href{https://dx.doi.org/10.1007/JHEP11(2010)004}{JHEP {\bfseries 11} (2010)
  004} {\ttfamily [\href{https://arxiv.org/abs/1008.2307}{arXiv:1008.2307}]}.

\bibitem{Endo:2015ixx}
M.~Endo, T.~Moroi, M.~M.~Nojiri, and Y.~Shoji, ``{Renormalization-Scale
  Uncertainty in the Decay Rate of False Vacuum},''
  \href{https://dx.doi.org/10.1007/JHEP01(2016)031}{JHEP {\bfseries 01} (2016)
  031} {\ttfamily [\href{https://arxiv.org/abs/1511.04860}{arXiv:1511.04860}]}.

\bibitem{Batell:2013bka}
B.~Batell, S.~Jung, and C.~E.~M.~Wagner, ``{Very Light Charginos and Higgs
  Decays},'' \href{https://dx.doi.org/10.1007/JHEP12(2013)075}{JHEP {\bfseries
  12} (2013) 075} {\ttfamily
  [\href{https://arxiv.org/abs/1309.2297}{arXiv:1309.2297}]}.

\bibitem{Endo:2014pja}
M.~Endo, T.~Kitahara, and T.~Yoshinaga, ``{Future Prospects for Stau in Higgs
  Coupling to Di-photon},''
  \href{https://dx.doi.org/10.1007/JHEP04(2014)139}{JHEP {\bfseries 04} (2014)
  139} {\ttfamily [\href{https://arxiv.org/abs/1401.3748}{arXiv:1401.3748}]}.

\bibitem{Aprile:2020vtw}
{\bfseries XENON} Collaboration, ``{Projected WIMP sensitivity of the XENONnT
  dark matter experiment},''
  \href{https://dx.doi.org/10.1088/1475-7516/2020/11/031}{JCAP {\bfseries 11}
  (2020) 031} {\ttfamily
  [\href{https://arxiv.org/abs/2007.08796}{arXiv:2007.08796}]}.

\bibitem{ATL-PHYS-PUB-2018-048}
{\bfseries ATLAS} Collaboration, ``{Prospects for searches for staus, charginos
  and neutralinos at the high luminosity LHC with the ATLAS Detector},''
  \href{https://cds.cern.ch/record/2651927}{ATL--PHYS--PUB--2018--048}, CERN,
  2018.

\bibitem{CidVidal:2018eel}
X.~Cid~Vidal, M.~D'Onofrio, P.~J.~Fox, R.~Torre, and K.~A.~Ulmer, eds.,
  ``Beyond the Standard Model physics at the HL-LHC and HE-LHC,''
  \href{https://dx.doi.org/10.23731/CYRM-2019-007.585}{CERN Yellow Reports:
  Monographs {\bfseries 7} (2019) 585--865} {\ttfamily
  [\href{https://arxiv.org/abs/1812.07831}{arXiv:1812.07831}]}.

\bibitem{Gori:2014oua}
S.~Gori, S.~Jung, L.-T.~Wang, and J.~D.~Wells, ``{Prospects for Electroweakino
  Discovery at a 100 TeV Hadron Collider},''
  \href{https://dx.doi.org/10.1007/JHEP12(2014)108}{JHEP {\bfseries 12} (2014)
  108}
{\ttfamily [\href{https://arxiv.org/abs/1410.6287}{arXiv:1410.6287}]}.
%%CITATION = ARXIV:1410.6287;%%.

\bibitem{Bramante:2014tba}
J.~Bramante, {\em et al.}, ``{Relic Neutralino Surface at a 100 TeV
  Collider},'' \href{https://dx.doi.org/10.1103/PhysRevD.91.054015}{Phys.\
  Rev.\  {\bfseries D91} (2015) 054015}
{\ttfamily [\href{https://arxiv.org/abs/1412.4789}{arXiv:1412.4789}]}.
%%CITATION = ARXIV:1412.4789;%%.

\bibitem{Matsumoto:2017vfu}
S.~Matsumoto, S.~Shirai, and M.~Takeuchi, ``{Indirect Probe of Electroweakly
  Interacting Particles at the High-Luminosity Large Hadron Collider},''
  \href{https://dx.doi.org/10.1007/JHEP06(2018)049}{JHEP {\bfseries 06} (2018)
  049}
{\ttfamily [\href{https://arxiv.org/abs/1711.05449}{arXiv:1711.05449}]}.
%%CITATION = ARXIV:1711.05449;%%.

\bibitem{Matsumoto:2018ioi}
S.~Matsumoto, S.~Shirai, and M.~Takeuchi, ``{Indirect Probe of
  Electroweak-Interacting Particles with Mono-Lepton Signatures at Hadron
  Colliders},'' \href{https://dx.doi.org/10.1007/JHEP03(2019)076}{JHEP
  {\bfseries 03} (2019) 076}
{\ttfamily [\href{https://arxiv.org/abs/1810.12234}{arXiv:1810.12234}]}.
%%CITATION = ARXIV:1810.12234;%%.

\bibitem{Chigusa:2018vxz}
S.~Chigusa, Y.~Ema, and T.~Moroi, ``{Probing electroweakly interacting massive
  particles with Drell–Yan process at 100 TeV hadron colliders},''
  \href{https://dx.doi.org/10.1016/j.physletb.2018.12.011}{Phys.\  Lett.\
  {\bfseries B789} (2019) 106--113}
{\ttfamily [\href{https://arxiv.org/abs/1810.07349}{arXiv:1810.07349}]}.
%%CITATION = ARXIV:1810.07349;%%.

\bibitem{Abe:2019egv}
T.~Abe, S.~Chigusa, Y.~Ema, and T.~Moroi, ``{Indirect studies of electroweakly
  interacting particles at 100 TeV hadron colliders},''
  \href{https://dx.doi.org/10.1103/PhysRevD.100.055018}{Phys.\  Rev.\
  {\bfseries D100} (2019) 055018}
{\ttfamily [\href{https://arxiv.org/abs/1904.11162}{arXiv:1904.11162}]}.
%%CITATION = ARXIV:1904.11162;%%.

\bibitem{Endo:2013xka}
M.~Endo, K.~Hamaguchi, S.~Iwamoto, T.~Kitahara, and T.~Moroi, ``{Reconstructing
  Supersymmetric Contribution to Muon Anomalous Magnetic Dipole Moment at
  ILC},'' \href{https://dx.doi.org/10.1016/j.physletb.2013.11.068}{Phys.\
  Lett.\  B {\bfseries 728} (2014) 274--281} {\ttfamily
  [\href{https://arxiv.org/abs/1310.4496}{arXiv:1310.4496}]}.

\bibitem{Beresford:2018pbt}
L.~Beresford and J.~Liu, ``{Search Strategy for Sleptons and Dark Matter Using
  the LHC as a Photon Collider},''
  \href{https://dx.doi.org/10.1103/PhysRevLett.123.141801}{Phys.\  Rev.\
  Lett.\  {\bfseries 123} (2019) 141801} {\ttfamily
  [\href{https://arxiv.org/abs/1811.06465}{arXiv:1811.06465}]}.

\end{thebibliography}\endgroup

\end{document}

% LocalWords:  Motoi Endo Koichi Hamaguchi Sho Iwamoto Teppei Kitahara
% LocalWords:  IPNS Tsukuba Ibaraki Sokendai Kavli WPI UTIAS Kashiwa
% LocalWords:  Bunkyo degli Studi Padova Marzolo Padua Technion Maskawa
% LocalWords:  INFN Sezione ELTE Kobayashi leptonic Drell Yan SLHA
% LocalWords:  Eq Eqs GeV TeV ATLAS CMS
% LocalWords:  slepton sleptons chargino charginos neutralino neutralinos
% LocalWords:  higgsino higgsinos
% LocalWords:  boson bosons supersymmetry hadronic photonic BNL BSM SUSY
% LocalWords:  Brookhaven CODATA Fermilab SUSY CP
% LocalWords:  electroweakino electroweakinos